\documentclass[final,pre,showpacs,amsmath,amssymb,longbibliography]{revtex4-1}

\usepackage{graphicx}
\usepackage{color}
\usepackage{dcolumn}
\usepackage{bm}
\usepackage{epstopdf}
\usepackage{hyperref}

	\usepackage{fancyheadings}
\renewcommand{\thispagestyle}[1]{} 

\begin{document}

\title{Phase Diagram of the $J_1$ - $J_2$ Frustrated Anisotropic Antiferromagnet with Spin $S=1$ on the Quadratic Lattice}

\author{T. Balcerzak}
\email{tadeusz.balcerzak@gmail.com}
\author{K. Sza{\l}owski}
\email{kszalowski@uni.lodz.pl}
\affiliation{%
Department of Solid State Physics, Faculty of Physics and Applied Informatics,\\
University of \L\'{o}d\'{z}, ulica Pomorska 149/153, 90-236 \L\'{o}d\'{z}, Poland
}%

\author{A. Bob\'{a}k}
\author{M. \v{Z}ukovi\v{c}}

\affiliation{%
Department of Theoretical Physics and Astrophysics, Faculty of Science,\\
P. J. \v{S}af\'arik University, Park Angelinum 9, 041 54 Ko\v{s}ice, Slovak Republic
}%

\date{\today}

\begin{abstract}
In the paper the phase diagram of $J_1-J_2$ frustrated antiferromagnet with spin $S=1$ and single-ion anisotropy is studied on the planar quadratic lattice in the cluster approximation. The Bogolyubov inequality is adopted for the Gibbs energy calculation for the case of $2 \times 2$ and $4 \times 4$ clusters. On this basis, the ranges of existence of the antiferromagnetic, superantiferromagnetic and paramagnetic phases are investigated for the antiferromagnetic nearest-neighbour ($J_1<0$) and next-nearest-neighbour ($J_2<0$) interactions. In particular, the occurrence of tricritical and triple points is discussed and a comparison between the results for  $2 \times 2$ and $4 \times 4$ clusters is made.  The results are also compared with the classical MFA method, adopted here for the model in question, as well as with selected literature results for particular choices of interaction parameters.
\end{abstract}

\maketitle

\section{Introduction}

In magnetic systems the spin frustration arises frequently due to the presence of competing exchange interactions. The exemplary theoretical system is the Ising spin $S=1/2$ model on the square lattice with the nearest- neighbour (NN) and next-nearest-neighbour (NNN) interactions (so called $J_1 - J_2$ model) \cite {Binder1980,Landau1980,Jin2013}. In this model the NN interactions, $J_1$, are either ferromagnetic or antiferromagnetic, whereas the NNN interactions, $J_2$, are of the antiferromagnetic character. The physical realisation of the Ising system have been the $\rm K_2 Co F_4$ and $\rm Rb_2 Co F_4$ compounds \cite{Samuelsen1974, Ikeda1974}, which are the quasi-two-dimensional anisotropic antiferromagnets with spin $S=1/2$. Another example is the $\rm FeCl_2$ compound \cite{Jongh1974}, which is the quasi-two-dimensional anisotropic ferromagnet with spin $S=1$. 

In theoretical studies of the $J_1 - J_2$ models many efforts have been focused on calculations of the phase diagrams, including the existence of ferromagnetic (F), antiferromagnetic (AF), superantiferromagnetic (SAF) and paramagnetic (P) phases \cite {Binder1980,Landau1980,Jin2013,Moran-Lopez1993,Bobak1996, dosAnjos2008,Kalz2008, Bobak2015,Murtazaev2014}. In the above-mentioned papers, the order of phase transitions and possible occurrence of the tricritical points have been widely discussed using various methods. Apart from the Ising model with spin $S=1/2$, the investigations include also the Heisenberg model with spin $S=1/2$ \cite{Richter2010}, the spin $S=1$ Blume-Capel model with transverse crystal field \cite{Viana2017}, as well as the Blume-Capel model with random field \cite{Santos2018}.

As far as the spin $S=1$ systems are concerned, the frustrated $J_1 - J_2$ model including both NN and NNN bilinear interactions, together with the single-ion anisotropy term, has been studied in Ref.\cite{Bobak1996}. 
In one particular case, when $J_2=0$, the model in question simplifies and becomes equivalent to the Blume-Capel model with NN interactions and single-ion anisotropy \cite{Blume1966,Capel1966}. It should be mentioned that in order to solve the simplified model many approximative methods have been applied, for instance the Molecular Field Approximation (MFA) \cite{Balcerzak1995}, Effective Field Theory (EFT) \cite{Fittipaldi1989}, Cluster Variational Method in Pair Approximation (CVMPA) \cite{Balcerzak1995}, or Monte Carlo (MC) technique \cite{Wilding1996,Xavier1998,Silva2006,Yuksel2009,Malakis2010,
Plascak2013,Ryu2013,Kwak2015,Zierenberg2017} (including a combined MC-MFA approach \cite{Viana2016}) as well as High Temperature Series Expansion (HTSE) approach \cite{Butera2018} or transfer matrix technique \cite{Jung2017} and other techniques \cite{Kim2014}. However, for the case when NNN interactions are present, the literature results are far less common, to mention the works in EFT approximation \cite{Bobak1996} or in MFA approximation \cite{Badehdah1998}.

The spin $S=1$ model, in which the NNN interactions are competing with NN interactions, and the single-ion anisotropy plays a significant role, is most interesting because it leads to multi-causal spin frustrations. At the same time it brings additional difficulties concerning theoretical treatment and, as a consequence, some methods cannot be adopted for its studies. For instance, adaptation of the CVMPA method for frustrated systems occurred to be rather difficult even for spin $S=1/2$ \cite{Balcerzak2014} and therefore that method would be even more challenging for spin $S=1$. This concerns especially the range of strong frustration, at low temperatures. However, one should mention that in some limited cases, for instance when the long-range RKKY interactions are present and the temperatures are high enough (being around the phase transition temperature), the CVMPA method may lead to the acceptable results \cite{Balcerzak2004}.

As regards the EFT method, its main deficiency lies in the fact that this approach does not yield directly the Gibbs energy, and therefore such a method is not useful for studies of the 1st order phase transitions. For this reason, in Ref.\cite{Bobak1996} the EFT approach was restricted only to investigations of continuous phase transitions, including the existence of the tricritical point. A novel idea of obtaining an analytical form of the Gibbs energy for spin $S=1$ has been presented in Ref.\cite{Costabile2012}. However, in our opinion that approach was based on the simplified \emph{ansatz} which might be somehow disputable. It should also be mentioned that a self-consistent method of obtaining the Gibbs energy within EFT has been proposed in Ref.\cite{Balcerzak2008}. Unfortunately, that approach was developed only for spin $S=1/2$ and has not been extended yet for higher spin values.

In the state of the art, the possible analytical approach to frustrated spin $S=1$ systems should be different from the methods mentioned above. 
The desired method would be universal regarding the treatment of various spin magnitudes $S$, and also would be controllable by means of systematic approximations.
One of the possible approach, especially convenient from the point of view of thermodynamic calculations (yielding the magnetic Gibbs energy), is based on the Gibbs-Bogolyubov inequality. Derivation of this general method has been presented is several classical papers \cite{Peierls1938,Feynman1955,Isihara1968,Falk1970} and its applications have been reviewed for example in Ref.~\cite{Kuzemsky2015}. The essence of the method lies in finding an upper bound for the actual Gibbs energy for the studied system by selecting a parameter-dependent trial Hamiltonian and solving it exactly. The parameter or parameters are determined by variational minimization of the the bound (right-hand side of the inequality) for the Gibbs energy. The selection of various trial Hamiltonians allows the systematic construction of a well-controlled wide class of approximations. By convention, the method has been used for the development of the single-particle classical MFA, where the ordinary self-consistent equation for magnetization is shown to minimize the Gibbs energy bound (for details see, for example, Refs.~\cite{Yeomans} and \cite{Strecka2015}). Recently, the extension of the method, for the case when the system can be divided into multi-spin identical clusters, has been demonstrated in Refs. \cite{Jin2013,Santos2017,Mendes2018}. 

In this approach, the cluster shape should reflect the symmetry of the lattice, and the number of spins is limited only by the computational capability, since the statistical sum inside the cluster is calculated numerically. The clusters with more rounded shape enable to increase the number of spin-spin bonds which are contained inside the clusters and are exactly taken into account. At the same time, for such clusters, the role of molecular fields acting on the cluster edges is minimised, which increases the accuracy of the method.

In the present paper we select this approach for studies of the frustrated $J_1 - J_2$ model with spin $S=1$ on the planar square lattice. Thus, we intend to fill the gap left in the investigations as a result of deficiencies of other analytical methods. In order to take into account the spin frustration, both the NN $(J_1)$ and NNN $(J_2)$ exchange interactions are assumed to be of the antiferromagnetic character, and the single-ion anisotropy term $(D)$ is simultaneously included.

Out aim is to investigate the phase diagram of the model in the $D - J_2$ space in the full range of temperatures $T \ge 0$. In particular, the existence of the tricritical points (TCPs), separating the continuous (2nd order) and discontinuous (1st order) phase transition boundary, will be studied. In the systematic approach the results will be obtained and compared for the cases of 1-site clusters (classical MFA), 4-site $(2 \times 2)$ clusters and 16-site $(4 \times 4)$ clusters. For some special cases of the Hamiltonian parameters, like $J_2=0$ (Blume-Capel model with NN interactions) or $D \to \infty$ (Ising model with NN and NNN interactions), the results will be compared with those existing in the literature and obtained using different methods.

\section{\label{sec:theory}Theoretical model}

The spin $S=1$ Hamiltonian is assumed in the form of:
\begin{equation}
{\mathcal H}=-J_1\sum_{<i,j>}S_i^z S_j^z -J_2\sum_{<<i,j>>}S_i^z S_j^z -D\sum_{i}\left(S_i^z\right)^2-h\sum_{i} S_i^z
\label{eq1}
\end{equation}
where $J_1<0$ and $J_2<0$ are the NN and NNN antiferromagnetic exchange interactions, respectively. $D$ is the single-ion anisotropy parameter and $h$ stands for the external magnetic field. $S_i^z$ denotes the $z$-component of the spin in $i$-th lattice site, and $S_i^z = \pm 1,0$.

The Gibbs free energy $G$ for the Hamiltonian ${\mathcal H}$ is found from the the Gibbs-Bogolyubov inequality \cite{Peierls1938,Feynman1955,Isihara1968,Falk1970}:
\begin{equation}
G \le G_0 + \left< {\mathcal H} - {\mathcal H}_0\right>_0
\label{eq2}
\end{equation}
assuming that $G$ reaches its upper limit. $G_0$ is the Gibbs free energy for the trial Hamiltonian ${\mathcal H}_0$ and the thermal averaging $\left< {\mathcal H} - {\mathcal H}_0 \right>_0$ is performed with the trial density matrix, namely:
\begin{equation}
\left< {\mathcal H} - {\mathcal H}_0 \right>_0 = \frac{ {\rm Tr} \,({\mathcal H} - {\mathcal H}_0) e^{- \beta {\mathcal H}_0}}{{\rm Tr} \, e^{- \beta {\mathcal H}_0}}.
\label{eq2a}
\end{equation} 

In general, $G_0$ can be written in the form:
\begin{equation}
G_0 = - k_{\rm B} T \ln \left( {\rm Tr} \, e^{- \beta {\mathcal H}_0} \right),
\label{eq3}
\end{equation}
where $\beta= 1/ k_{\rm B} T$.

We assume that the total number of spins (lattice sites) in the system is denoted by $N$. If the system can be divided into $n$-atomic $(n>1)$ identical and mutually exclusive clusters $c$, the total trial Hamiltonian can be proposed as a sum of the cluster trial Hamiltonians ${\mathcal H}_0^c$:
\begin{equation}
{\mathcal H}_0 =  \sum_{c}^{N/n}{\mathcal H}_0^c .
\label{eq4}
\end{equation}
Then, the Gibbs energy per lattice site is given by:
\begin{equation}
\frac{G}{N} = - \frac{1}{n}k_{\rm B} T \ln \left( {\rm Tr}_c \, e^{- \beta {\mathcal H}_0^c} \right)
+ \frac{1}{N}\left< {\mathcal H} - {\mathcal H}_0\right>_0 ,
\label{eq5}
\end{equation}
where ${\rm Tr}_c$ denotes the trace taken over the cluster $c$.
In this paper we assume two sizes of the clusters reflecting the magnetic symmetry of the AF and SAF phases:  $2 \times 2$ clusters with $n=4$,  and $4 \times 4$ clusters with $n=16$. Illustrations of these clusters are presented in Fig.~\ref{F1}, both for AF and SAF phase, in the two-sublattice model. Solid and dashed lines correspond to NN ($J_1$) and NNN  ($J_2$) interactions, respectively. These interactions, being of intracluster type, are identical with the interactions forming the original Hamiltonian ${\mathcal H}$. The edge spins, interacting with the molecular fields, are numbered by $i$-position $(i=1,2,3, \ldots)$ in accordance with the spins notations $S_i$ in corresponding formulas (see Appendices \ref{sec:appendixA}-\ref{sec:appendixC}).

The cluster trial Hamiltonian can be decoupled as:
\begin{equation}
{\mathcal H}_0^c =  \bar{{\mathcal H}}_0 +  {\mathcal H}_0^{\prime}.
\label{eq6}
\end{equation}
$\bar{{\mathcal H}}_0$ describes all NN and NNN interactions of the spins inside the cluster. These interactions (marked graphically by solid and dashed lines in Fig.~\ref{F1}) are taken into account exactly, what can be done even for large $n$ if the corresponding traces in Eqs. \ref{eq2a} and \ref{eq5} are performed numerically. Besides, ${\mathcal H}_0^{\prime}$ describes interaction of the edge spins with the molecular fields acting on these sites from the cluster neighbourhood. The molecular field parameters are introduced as an approximation, which is necessary for factorization of the trial partition function ${\rm Tr} \, e^{- \beta {\mathcal H}_0}$. They can be determined from the minimisation condition of the total Gibbs energy $G$ with respect to these parameters. It should be noted that the molecular fields depend on the spin location at the cluster edge, as well as on the sublattice index $a$ or $b$, as the (super)antiferromagnetic system is divided in such a way. The derivation of the appropriate formulas for the variational parameters is, in general, analogous to the case of ordinary MFA (for details see, for example, Refs.~\cite{Yeomans} and \cite{Strecka2015}).

In Appendices \ref{sec:appendixA}-\ref{sec:appendixC} we collected the corresponding formulas for ${\mathcal H}_0^{\prime}$, the molecular fields parameters in equilibrium, the perturbative terms $\frac{1}{N}\left< {\mathcal H} - {\mathcal H}_0\right>_0$ in Eq.\ref{eq5}, as well as for the local magnetizations at the cluster edges, which are determined self-consistently. All those formulas are presented for $2 \times 2$ and $4 \times 4$ clusters, both for AF and SAF phases. For completeness, the relationships in the classical MFA, based on the 1-site clusters for AF and SAF phases, are also included there.

From the formulas presented in the Appendices \ref{sec:appendixA}-\ref{sec:appendixC}, the Gibbs free energies of AF and SAF phases can be calculated numerically both for the classical MFA, as well as for the cluster method with $2 \times 2$ and $4 \times 4$ clusters, on the basis of the general Eq.(\ref{eq5}). In turn, the Gibbs potential of the paramagnetic phase is obtained from the same formulas only by setting the molecular field parameters equal to zero.
After the Gibbs free energies are determined, all the thermodynamic properties can be  calculated self-consistently. In particular, for the phase diagrams determination, the Gibbs energies per lattice site, $G/N$, (i.e., the chemical potentials) of coexisting phases in equilibrium should be equated, for the same temperature $T$ and external field $h$.

The numerical calculations of the phase diagrams, based on the presented formalism, will be shown and discussed in the next Section \ref{sec:results} for the spontaneously ordered phases, i.e., when the external magnetic field is absent $(h=0)$.

\section{\label{sec:results}Numerical results and discussion}

The numerical results are presented starting from the ground-state phase diagram (Fig.~\ref{F2}) in the $D/|J_1|$ - $J_2/|J_1|$ space. The areas corresponding to the antiferomagnetic, super-antiferromagnetic and paramagnetic (spin-zero nonmagnetic) phases are denoted by AF, SAF and P, respectively. The spin states in these phases are  depicted schematically. It can be noted that the diagram for $T=0$ is analogous to that presented in Ref.\cite{Bobak1996}. However, the ferromagnetic phase obtained in  Ref.\cite{Bobak1996} is replaced here by the antiferromagnetic one. Moreover, we found a new mixed phase existing solely on the P/SAF line (and schematically depicted in Fig.~\ref{F2}) which has not been reported previously. In this new phase the spin-zero state is fixed at every second spin. In such a case, the NN interactions do not contribute to the magnetic energy, while the NNN interactions cancel out with the single-ion anisotropy terms, if $D/|J_1| = 2J_2/|J_1|$. Therefore, the energy of this phase is zero in the ground state, the same as for the paramagnetic (spin-zero nonmagnetic) phase. The mixed state is stable only in the ground state (for $T=0$). It is worth noticing that the SAF state presented in Fig.~\ref{F2} is the same as the "stripped" state according to the Fig.1 of Ref.\cite{Jin2013}. We also would like to stress the point that the ground state phase diagram presented in Fig.~\ref{F2} is exact, not depending on the cluster size or any approximation. The symmetry of the ground states is connected with the symmetry of the underlying square lattice. It has also been found that the phase transitions between all the ground states are discontinuous (of the 1st order).

The order parameters for each ground state are the sublattice magnetizations, which are characteristic of a given cluster. For instance, for $4 \times 4$ cluster the edge magnetizations for AF phase are given by the set of 4 Eqs.\ref{a18}, whereas for SAF phase the set of 6 Eqs.\ref{a22} is appropriate. Having solved these two sets of equations we find the Gibbs energies (Eq.\ref{eq5}) associated with them. Then, the stable phase is chosen, corresponding to that solution for which the Gibbs energy reaches the lowest value.
The criterion based on the Gibbs potential minimization allows the determination of the stability regions for AF, SAF and P phases, not only in the ground state (where it gives exact result), but also in the finite-temperature phase diagrams.

In Fig.~\ref{F3} the finite-temperature phase diagram is presented in $D/|J_1|$ - $k_{\rm B}T/|J_1|$ coordinates for the absence of NNN interaction ($J_2=0$). In this case the $J_1$ - $J_2$ model reduces to the pure Blume-Capel model which, for the ferromagnetic case, has been intensively studied in literature. In particular, for $D=0$ the Curie temperatures, $k_{\rm B}T_{\rm C}/|J_1|$, have been found by different methods, giving the following results: 2.667 (MFA \cite{Balcerzak1995}), 2.322 \cite{SaBarreto2012}, 2.220 (present, $4 \times 4$ cluster), 2.188 (EFT \cite{Fittipaldi1989}) and 2.066 (CVMPA \cite{Balcerzak1995}) or 1.952 \cite{Akinci2012}. According to our knowledge, the best results for $D=0$ were obtained either by LTSE method (namely $k_{\rm B}T_{\rm C}/|J_1|=1.6936$ \cite{Enting1994}), HTSE-LTSE method: 1.69378 \cite{Butera2018} or using MC approach: 1.690 \cite{Yuksel2009}, 1.681 \cite{Xavier1998}, 1.695 \cite{Beale1986}. Also the HTSE-LTSE results for $D/|J_1|=-0.5$ equal to 1.5664, for $D/|J_1|=-1.0$ equal to 1.3986, for $D/|J_1|=-1.5$ equal to 1.1467 and for $D/|J_1|=-1.9$ equal to 0.766 (all after Ref.~\cite{Butera2018}) can be mentioned.

On the other hand, for $J_2=0$ and $D/|J_1|<0$ the model exhibits existence of the tricritical point (TCP) where the 2nd order phase transition lines (denoted by continuous lines in Fig.~\ref{F3}), separating antiferromagnetic and paramagnetic phases, become the 1st order phase transitions (represented by the dashed lines). The TCPs positions are marked by the bold dots in Fig.~\ref{F3}.
The coordinates of tricritical points for the antiferromagnetic model, with the assumption that the N\'{e}el temperature $T_{\rm N}$ for $J_1 <0$ equates with the Curie temperature $T_{\rm C}$ for $J_1 >0$, are collected in Table I for various methods.

From the comparison of numerical results presented in Fig.~\ref{F3} and Table \ref{table1} one could conclude that the present approximation based on $4 \times 4$ cluster for $J_2=0$, i.e., without frustration, is more accurate than the classical MFA method (and better than approximation based on $2 \times 2$ cluster) but is less accurate than Cluster Variational Method in Pair Approximation (CVMPA). The accuracy obtained here for  $4 \times 4$ cluster in the case of $J_2=0$, as regards $k_{\rm B}T_{\rm N}/|J_1|$ for $D=0$ and the position of TCP, is comparable with accuracy of the Effective Field Theory (EFT). It is also worth noticing that for $T \to 0$ all the curves give convergent results and they end at $D/|J_1|=-2$, in agreement with the ground-state phase diagram for $J_2/|J_1|=0$ (Fig.~\ref{F2}).

\begin{table*}[ht]
\caption{\label{table1}Coordinates of TCP for the pure Blume-Capel model, with NN interactions only, obtained by different methods.}
\vspace {5 mm}
\begin{tabular}{|c||c|l|}
\hline {method} & {$D/|J_1|$} & {$k_{\rm B}T_{N}^{*}/|J_1|$}\\ 
\hline
\hline {MFA \cite{Balcerzak1995}} & { $-1.848$} & {$\;\;\; 1.333$}  \\ 
\hline {present ($4 \times 4$ cluster)} & { $-1.878$} & {$\;\;\; 1.020$} \\ 
\hline {EFT \cite{Fittipaldi1989}} & { $-1.880$} & {$\;\;\; 1.01$} \\ 
\hline {CVMPA \cite{Balcerzak1995}} & { $-1.900$} & {$\;\;\; 0.913$}  \\
\hline {EFT \cite{Akinci2012}} & { $-1.912$} & {$\;\;\; 0.846$}  \\
\hline {MFT-MC \cite{Viana2016}} & { $-1.941$} & {$\;\;\; 0.725$}  \\
\hline {MC \cite{Xavier1998}} & { $-1.965$} & {$\;\;\; 0.609$} \\ 
\hline {MC \cite{Silva2006}} & { $-1.974$} & {$\;\;\; 0.56$} \\ 
\hline {transfer matrix \cite{Kwak2015}} & { $-1.9660$} & {$\;\;\; 0.6080$} \\ 
\hline {MC \cite{Jung2017}} & { $-1.96582$} & {$\;\;\; 0.60858$} \\ 

\hline
\end{tabular}
\end{table*}

According to the discussion in Introduction, the methods like CVPMA and EFT are not suitable to be fully applied to studies of the frustrated model, i.e., when $J_2/|J_1|< 0$, in the whole range of Hamiltonian parameters and arbitrary temperature. Therefore, in further investigations of the $J_1$ - $J_2$ model with spin $S=1$ and single-ion anisotropy, we decided to use the cluster approach, which formalism has been described in the Section \ref{sec:theory}. In numerical calculations we start from the smallest possible cluster, i.e., consisting of a single atom, which corresponds to the classical MFA method. As far as we know, the MFA method has not been exploited yet with the present model, therefore such studies are justified and purposeful as a first step. The results of MFA calculations for $J_1<0$, $J_2 \le 0$ and $D \le 0$ are collected in Figs.\ref{F4}-\ref{F6}.

In Fig.~\ref{F4} the MFA phase diagram is shown illustrating the dimensionless N\'{e}el temperature $k_{\rm B}T_{N}/|J_1|$ of AF phase vs. anisotropy parameter $D/|J_1|$. Various $J_2/|J_1|$ ratios from the range of $0 \le J_2/|J_1| \le -0.5$ are chosen. As before, the solid lines present the continuous (2nd order) phase transitions, while the dashed lines correspond to the discontinuous (1st order) ones. The evolution of TCPs (marked by the bold dots) is seen when the ratio $J_2/|J_1|$ changes. It can be concluded that the TCP position moves linearly with $J_2/|J_1|$ change and the temperature
$k_{\rm B}T_{N}^{*}/|J_1|$ decreases when $J_2/|J_1|$ is shifted from 0 to -0.5. At the same time, for $T \to 0$ the range of AF phase in $D$-direction decreases linearly with diminishing of $J_2/|J_1|$ ratio, in agreement with the ground-state phase diagram (Fig.~\ref{F2}).

Fig.~\ref{F5} presents a continuation of the phase diagram from Fig.~\ref{F4} for the range of $J_2/|J_1| <-0.5$. In this case the AF phase is replaced by the SAF one. The evolution of TCPs for SAF phase proceeds in inverse direction, i.e., for decreasing $J_2/|J_1|$ ratio the TCP temperature $k_{\rm B}T_{N}^{*}/|J_1|$ increases linearly. As a result, the range of SAF phase expands in $D$-direction simultaneously with evolution of TCPs, which is expected from the ground-state phase diagram (Fig.~\ref{F2}).

It is also interesting to present the MFA phase diagram in $k_{\rm B}T/|J_1|$ - $J_2/|J_1|$ coordinates, when the $D/|J_1|$ parameters are constant. Such diagram is illustrated in Fig.~\ref{F6}. It can be noted that the curves are symmetric with respect to $J_2/|J_1| = -0.5$, which separates the ranges of AF and SAF phases. For the $D/|J_1| = 0$ and $D/|J_1| = -0.5$ curves, the phase transitions are only of the 2nd order and the three phases: AF, SAF and P meet at the triple point, which is located at $J_2/|J_1| = -0.5$. It is interesting that for $D/|J_1| = 0$, the diagram is linear and the temperature of the triple point at $J_2/|J_1| = -0.5$ amounts exactly to $k_{\rm B}T_{N}/|J_1| = 1 \frac{1}{3}$, which is half of the value $k_{\rm B}T_{N}/|J_1| = 2 \frac{2}{3}$ obtained for $J_2/|J_1| = 0$ and $J_2/|J_1| = -1$. For $D/|J_1| = -1$, the TCPs occur symmetrically for the AF and SAF phases and the triple point is moved to $T=0$. For $D/|J_1| < -1$, the temperatures of TCP points tend to increase symmetrically with respect to $J_2/|J_1| = -0.5$, whereas the AF and SAF phases become separated, leaving the space between them for the paramagnetic phase. Again, this behaviour is in accordance with the ground-state phase diagram (Fig.~\ref{F2}).

Having calculated the classical MFA phase diagram, as a next step we demonstrate the improvement which can result from considering larger clusters. The results are illustrated in Figs.\ref{F7}-\ref{F9}, which are prepared in the same coordinates as Figs.\ref{F4}-\ref{F6}, respectively. For instance, in Fig.~\ref{F7} the N\'{e}el temperature of AF phase, $k_{\rm B}T_{N}/|J_1|$, is plotted as a function of the single-ion anisotropy, $D/|J_1|$, for three values of the NNN exchange interaction parameters:  $J_2/|J_1| = 0$, -0.25 and -0.5. Thin lines correspond to the smaller  clusters ($2 \times 2$) whereas the thick lines are for the larger clusters ($4 \times 4$). As before, solid lines denote the continuous phase transitions, while the dashed lines are for 1st order ones. The positions of TCPs are marked by the bold dots. By comparison of Fig.~\ref{F7} with Fig.~\ref{F4} one can conclude that the cluster phase diagram is qualitatively similar to that in MFA. However, comparing it quantitatively, an increase of the cluster size results in reducing the N\'{e}el temperatures and lowering of the position of TCPs. We note that the changes in TCP temperatures are much more evident than the changes in $D/|J_1|$ coordinates for these points. On the other hand, in the low-temperature range, the differences between curves obtained for different clusters become negligible, and for $T \to 0$ both curves tend to the same point resulting also from the ground-state phase diagram.

By the same token, Fig.~\ref{F8} can be compared with Fig.~\ref{F5}. In Fig.~\ref{F8} the N\'{e}el temperatures, corresponding to SAF phase, are plotted vs. anisotropy $D/|J_1|$, for the three values of NNN exchange interactions:  $J_2/|J_1| = -0.6$, -0.8 and -1.0. The convention regarding thickness and style of the lines remains the same as in previous figures. In this case an interesting feature is found for the curve with $J_2/|J_1| = -0.6$ and $2 \times 2$ cluster: namely, the existence of two tricritical points. On this basis, it could be concluded that for this curve the 2nd order transitions are confined to the region between two TCPs. However, on the curve prepared for $4 \times 4$ clusters, and the same value of $J_2/|J_1| = -0.6$, this phenomenon is not confirmed, since only one TCP exists there. Noting that the diagram for $4 \times 4$ clusters is qualitatively similar to the MFA diagram, we suppose that the existence of two TCPs on the curve with $J_2/|J_1| = -0.6$ is rather an artefact of approximation connected with this specific $2 \times 2$ cluster size. The origin of the 1st order transitions appearing on the upper part of the line with $J_2/|J_1| = -0.6$ for $2 \times 2$ cluster will be further commented on the basis of the next figure (Fig.~\ref{F9}).

In Fig.~\ref{F9} the N\'{e}el temperatures, corresponding both to AF and SAF phases, are plotted vs. NNN exchange integral $J_2/|J_1|$ for the three values of single-ion anisotropy: $D/|J_1| = 0$, -1.0 and -1.5. This diagram, prepared for $2 \times 2$ and $4 \times 4$ clusters, can be compared to Fig.~\ref{F6} prepared for MFA. The convention regarding thickness and style of the lines remains unchanged. First of all, it should be noted that the ideal symmetry of the curves presented in Fig.~\ref{F6} is broken when larger clusters are taken into account. This is most clearly seen in the vicinity of $J_2/|J_1| = -0.5$, where, for instance, the vertical dashed lines, for $D/|J_1| = 0$ and low temperatures, are bended towards lower values of $J_2/|J_1|$ when $T$ increases. As a result, the positions of triple points are shifted towards $J_2/|J_1| < -0.5$. This feature is interesting from the point of view of thermal behaviour of the system, since in the close vicinity of $J_2/|J_1| = -0.5$ (and $J_2/|J_1| < -0.5$) the SAF phase, existing for low temperatures, is replaced by the AF phase when the temperature increases, (and then the continuous phase transition to P phase takes place). The temperature transition from SAF to AF phase is then of the 1st order. Thus, Fig.~\ref{F9} shows that the range of existence of AF phase can exceed the value of $J_2/|J_1| = -0.5$, provided that the anisotropy $D$ is small and the temperature approaches the phase transition temperature.

Comparing the curves for $2 \times 2$ and $4 \times 4$ clusters in Fig.~\ref{F9} we see that the most pronounced changes occur for $D/|J_1| = 0$. In this case, for $4 \times 4$ cluster, the TCP position occurs at $J_2/|J_1| \approx -0.513$ and is the same as the triple point. However, for $2 \times 2$ cluster, the TCP position is shifted towards the value of $J_2/|J_1| \approx -0.609$, which is much lower than the triple point. It can be noted that for the point with coordinate $J_2/|J_1| = -0.6$ on the abscissa, and $D/|J_1| = 0$, the phase transition is of the 1st order. This is in accordance with Fig.~\ref{F8}, where the curve with the same parameters exhibited discontinuous phase transitions in the high-temperature regime, and the second TCP occurred as a consequence.
It can also be noticed that for stronger anisotropy, $D/|J_1| \le -1$, the phase diagram in Fig.~\ref{F9} is qualitatively similar to Fig.~\ref{F6}. However, as mentioned above, the symmetry observed previously with respect to $J_2/|J_1| = -0.5$ is not conserved in this case, and the values of N\'{e}el temperatures are much lower. It is also worth noticing that for $D/|J_1| = 0$, the phase transitions from AF to P phase are continuous in the range from $J_2/|J_1| = 0$ down to the triple point. This result is common both for classical MFA (Fig.~\ref{F6}) and the cluster approximation (Fig.~\ref{F9}). In this context the paper Ref.\cite{Bobak1996} should be mentioned, where the ferromagnetic Blume-Capel model with NN $J_1 >0$ and NNN $J_2 < 0$ interactions has been considered in EFT approximation. In that paper the TCP has been found for the ferromagnetic phase for $D/J_1 = 0$ in the range of $-0.5 < J_2/J_1 < 0$. Taking into account that by changing the sign of NN interactions from $J_1 >0$ to $J_1 < 0$ the ferromagnetic phase should be replaced by the antiferromagnetic one, lack of TCP in the present method illustrates the discrepancy between the results of EFT and the cluster approximation. In such situation, in order to clarify the problem, the MC simulations might be helpful.

To have an additional test of the present method we consider the case when  $D/|J_1| \to \infty$. In this limiting case the zero spin states are eliminated, and only $S^z = \pm 1$ states remain, which makes the present model equivalent to the Ising model with spin $S=1/2$. The spin $S=1/2$ Ising model on the square lattice with the ferromagnetic NN and aniferromagnetic NNN interactions has been studied in Ref.\cite{Jin2013}. Among several methods used there the cluster method has been exploited for $2 \times 2$ and $4 \times 4$ clusters, and the phase diagram has been obtained (see Fig.4 in Ref.\cite{Jin2013}). For comparison, our phase diagram is presented in Fig.~\ref{F10} for the value of $D/|J_1| = 100$, i.e., when the single-ion anisotropy is strong enough to approximate the present $S=1$ model by the Ising one. In Fig.~\ref{F10}, apart from the curves for $2 \times 2$ and $4 \times 4$ clusters, the classical MFA result is also presented for comparison. Again, the MFA diagram is symmetric with respect to $J_2/|J_1| = -0.5$, moreover, the triple point temperature amounts to $k_{\rm B}T_{N}/|J_1| = 2$, which is exactly half of the value obtained for $J_2/|J_1| = 0$ and  $J_2/|J_1| = -1$. No sign of the 1st order phase transitions is seen in the MFA diagram. In turn, for the cluster diagrams the symmetry is broken and the TCPs are found at $J_2/|J_1| = -0.659$ for the smaller cluster ($2 \times 2$) and at $J_2/|J_1| = -0.662$ for the larger one ($4 \times 4$). Both phase diagrams are very similar to those in Ref.\cite{Jin2013}, as far as the SAF phase is concerned. For instance, the analogous TCP has been found in Ref.\cite{Jin2013} for  $J_2/J_1 \approx 0.66$, which corresponds quite accurately to our TCP coordinate (taking into consideration the opposite sign convention in the Hamiltonian). It has been stated in Ref.\cite{Jin2013} that such value is very close to the result $J_2/J_1 \approx 0.67$ obtained in MC simulations. On the other hand, for this value of $J_2/J_1$, the model presented in Ref.\cite{Jin2013} belongs to the same universality class as the four-state Potts model. Moreover, for $J_2/J_1 \ge 0.67$ the Hamiltonian can be mapped onto Ashkin-Teller model, for which the phase transitions are of the second order. Therefore, the Potts point at $J_2/J_1 \approx 0.67$ seems to be well established as TCP for the model in question, and our testing calculations reproduce well the numerical result of Ref.\cite{Jin2013}.

It can also be noted that in Ref.\cite{Jin2013}, an additional TCP has been found, existing for ferromagnetic phase, however, only when $4 \times 4$ cluster was considered. It has been said there that the existence of this TCP is controversial since it has not been confirmed by MC simulations up to now. It should be stated that  the existence of analogous TCP in AF phase also cannot be confirmed from our phase diagram (Fig.~\ref{F10}), since in both approximations (corresponding to $2 \times 2$ and $4 \times 4$ clusters) only the continuous phase transitions were observed over the whole AF/P phase line. It is possible that the discrepancy between our result and the paper Ref.\cite{Jin2013} in this point is due to the details of the numerical procedure adopted. Anyway, the possibility of existence of the TCP in AF phase is still worth of further study, for instance, by employing the MC method.

In order to illustrate better the phase diagram from Fig.~\ref{F10}, in Fig.~\ref{F11} we plot the sublattice magnetizations $m_a$ and $m_b$ vs. dimensionless temperature $k_{\rm B}T/|J_1|$ for $2 \times 2$ cluster. The curves are plotted for $J_2/|J_1|=-0.49$, -0.6 and -0.7. The case of $J_2/|J_1|=-0.49$ corresponds to AF phase in Fig.~\ref{F10}, where the continuous phase transition takes place at $k_{\rm B}T_{N}/|J_1|=1.291$. On the other hand, for $J_2/|J_1|=-0.6$ and -0.7 the sublattice magnetizations of SAF phase are shown. Then, the discontinuous and continuous phase transition temperatures amount to $k_{\rm B}T_{N}/|J_1|=1.641$ and 2.179, respectively. The parameters were selected so that the phase transitions shown correspond to both sides of TCP for SAF phase, as seen in Fig.~\ref{F10}.

\section{Final remarks and conclusions}

In the paper the phase diagram of $J_1-J_2$ frustrated antiferromagnet with spin $S=1$ and single-ion anisotropy has been studied in the cluster approximation. For this purpose, the Bogolyubov inequality for the Gibbs energy has been adopted for $2 \times 2$ and $4 \times 4$ clusters. The results have been compared with those of the classical MFA method, as well as with some outcomes of various methods reported in the literature for the particular cases of interaction parameters.

The phase diagram has been comprehensively studied in the domain $J_1 <0$ and $J_2 <0$, in the presence of single-ion anisotropy $D$ (where, due to spin frustration, the AF, SAF and P phases play a dominant role). Moreover, in the ground state phase diagram, the existence of a new mixed phase along the SAF/P line was found. In the finite-temperature phase diagrams, the existence of tricritical and triple points have been thoroughly examined. In particular, it has been demonstrated that for $2 \times 2$ and $4 \times 4$ clusters, the position of the triple point, connecting AF, SAF and P phases, is slightly shifted towards $J_2/|J_1| < -0.5$. This contrasts with the MFA phase diagram, which was symmetric with respect to $J_2/|J_1| = -0.5$.

It can be concluded that the present method is suitable for the studies of frustrated systems, especially when taking into account the deficiencies of such approaches like EFT and CVMPA. The main advantage of the present method is that the Gibbs energy is obtained directly, which enables reliable search for the 1st order phase transition boundary in the phase diagrams. Moreover, when the large clusters are considered, a great part of spin-spin interactions in the system is included exactly, which increases the accuracy of calculations. However, the cluster size is limited in this method by the computational capability, since the computing time increases exponentially with the number of spins. For instance, in our case of spin $S=1$ and  $4 \times 4$ cluster, the  calculation of statistical sum involves $3^{16}$ states. When the iterative procedure is adopted for solving  equations for the molecular field parameters, such statistical sum must be recalculated many times in order to obtain the self-consistent solutions. The iterative procedure is necessary, because the cluster Hamiltonian $\mathcal H_0^c$ contains not only the spin variables but also the magnetizations (i.e., the mean values) of the edge spins. These magnetizations are contained in the part $\mathcal H_0^{\prime}$ of $\mathcal H_0^c$ (see Eq.\ref{eq6}) and must be re-calculated repeatedly with all the spin variables contained in the cluster. This fact makes the consideration of larger clusters challenging, provided that the shape of these clusters has to reproduce the symmetry of the magnetic phases.

It is seen from previous Section \ref{sec:results} that some differences between the results for $2 \times 2$ and $4 \times 4$ clusters, as well as with selected literature data, can occur. In particular, some controversy remains about the presence or absence of TCP for AF phase when $D=0$, which has been discussed in the context of EFT approximation \cite{Bobak1996}. The existence of additional TCP for AF phase, when the Ising limit ($D/|J_1| \to \infty$) is considered, is also disputable in the context of Ref.\cite{Jin2013}. In order to clarify these controversial points, application of other independent methods would be very welcome. For instance, we believe that our results may serve as a sound motivation for extensive MC or series-expansion-based studies of the system in question.

It should also be concluded that after obtaining the Gibbs energy, not only the phase diagrams, but also all the thermodynamic properties can be calculated. For instance, this includes thermal dependencies of the magnetization, correlation functions, the magnetic susceptibility, entropy and specific heat. During such calculations the temperature behaviour of the Gibbs energy and its derivatives is fully correct from the physical point of view. In particular, the Gibbs energy is always a concave function, decreasing with temperature. As the result, the entropy (which can be calculated as $S=-\left(\partial G/\partial T\right)_h$) is a positive function increasing with temperature. In particular, for $T=0$ the entropy amounts to $S=0$ for each phase, since the Gibbs energy becomes a constant function vs. temperature when $T \to 0$. On the other hand, for $T\to \infty$ entropy reaches the paramagnetic limit of $S/N=k_{\rm B} \ln{3}$, since the Gibbs energy is linear there. Increasing entropy as a function of the temperature guarantees that the magnetic specific heat is positive everywhere, which means the thermal stability of the system. The specific heat can be calculated as $C_h=T\left(\partial S/\partial T\right)_h=-T\left(\partial^2 G/\partial T^2\right)_h$ and is characterized by a peak at the phase transition temperature. Since the phase transition temperature depends on the frustration parameter $J_2$, one can expect that frustration markedly influences all the thermodynamic properties mentioned above, and the most visible changes should occur at the phase transition. Taking this into account, we would like to stress the point that the thermodynamic description of the model with frustration is constructed completely and self-consistently, basing on the physically correct behaviour of the Gibbs energy. However, the presentation of all the additional thermodynamic properties would exceed the frame of the present paper.

The application of the method to other systems with various underlying crystalline lattices and magnetic phases is possible. Also the magnets with inhomogeneous structure, for instance, diluted spin systems and systems with possible spin glass behaviour \cite{Schmidt2015}, can be studied. As a future application of the method for spin $S=1$ system, also the investigations of the frustrated Blume-Emery-Griffiths model, containing biquadratic interaction and single-ion anisotropy terms in addition to the NN and NNN bilinear interactions, might be of interest.\\

\appendix
\section{\label{sec:appendixA}Classical MFA}

The Gibbs free energies in the classical single-atom cluster approximation are given by the following formulas for AF and SAF phases: 
\begin{equation}
\frac{1}{N} G^{\rm AF}=2 J_1 m_a m_b + J_2 \left( m_a^2 + m_b^2 \right)- \frac{1}{2}k_{\rm B}T \left( \ln Z_a + \ln Z_b \right)
\label{a1}
\end{equation}
and
\begin{equation}
\frac{1}{N} G^{\rm SAF}=J_1 m_a m_b + \frac{1}{2} J_1 \left( m_a^2 + m_b^2 \right) + 2 J_2 m_a m_b- \frac{1}{2}k_{\rm B}T \left( \ln Z_a + \ln Z_b \right),
\label{a2}
\end{equation}
respectively.
The single-atom cluster statistical sums $Z_{\alpha}$, for sublattice $\alpha=a,b$ , are then given by:
\begin{equation}
Z_{\alpha}=2 e^{\beta D} \cosh \left[ \beta \left(\lambda_{\alpha}+ h \right) \right] +1,                                                                                                  \label{a3}
\end{equation}
 where $ \beta=\frac{1}{k_{\rm B}T}$,
and the sublattice magnetizations $m_{\alpha}=\left< S_{\alpha}^z \right>_0$ are of the form:
\begin{equation}
m_{\alpha}=\frac{2}{Z_{\alpha}} e^{\beta D} \sinh \left[ \beta \left(\lambda_{\alpha}+ h \right) \right].                                                                                          \label{a4}
\end{equation}
 
The molecular fields $\lambda_{\alpha}$ which minimize the Gibbs energies are presented as:\\
For AF phase:
\begin{eqnarray}
\lambda_a&=&4J_1m_b+4J_2m_a \nonumber \\
\lambda_b&=&4J_1m_a+4J_2m_b,
\label{a5}
\end{eqnarray}
and for SAF phase:
\begin{eqnarray}
\lambda_a&=&2J_1 \left(m_a+m_b \right)+4J_2m_b \nonumber \\
\lambda_b&=&2J_1 \left(m_a+m_b \right)+4J_2m_a.
\label{a6}
\end{eqnarray}
It can be shown that with the molecular fields presented above, the self-consistent equations for the magnetizations $m_{\alpha}$ (\ref{a4}) are equivalent to the necessary minimum conditions for the Gibbs energies:
\begin{equation}
\frac{\partial G^{\rm AF/SAF}}{\partial m_{\alpha}}=0,
\label{a6a}
\end{equation}
where $\alpha = a,b$, and $G^{\rm AF/SAF}$ are given by Eqs. (\ref{a1}) or (\ref{a4}), respectively.

\section{\label{sec:appendixB}$2 \times 2$ cluster}

\underline
{AF - phase:} (see Fig.~\ref{F1} (a))\\

Hamiltonian of the cluster boundary is of the form:
\begin{equation}
{\mathcal H}_0^{\prime} = -\left(S_1+S_4 \right) \lambda_a -\left(S_2+S_3 \right) \lambda_b,
\label{a7}
\end{equation}
where the molecular field parameters which minimize the Gibbs energy are given by:
\begin{eqnarray}
\lambda_a&=&2J_1m_b+3J_2m_a \nonumber \\
\lambda_b&=&2J_1m_a+3J_2m_b.
\label{a8}
\end{eqnarray}
The perturbative term in the Gibbs energy is presented as:
\begin{equation}
\frac{1}{N} \left< {\mathcal H} - {\mathcal H}_0 \right>_0 =J_1 m_a m_b + \frac{3}{4} J_2 \left( m_a^2 + m_b^2 \right).
\label{a9}
\end{equation} 
The boundary magnetizations are calculated self-consistently with the cluster trial Hamiltonian ${\mathcal H}_0^c$ from the necessary minimum conditions for the Gibbs energy, namely:
\begin{equation}
\frac{\partial G}{\partial m_{\alpha}}=0,
\label{a9a}
\end{equation}
where $\alpha = a,b$, and $G$ is given by Eq. (\ref{eq5}). From Eqs. (\ref{a9a}) we obtain:

\begin{eqnarray}
m_a&=&\frac{1}{2} \left< S_1+S_4 \right>_0 \nonumber \\
m_b&=&\frac{1}{2} \left< S_2+S_3 \right>_0.
\label{a10}
\end{eqnarray}

\underline
{SAF - phase:} (see Fig.~\ref{F1} (b))\\

Hamiltonian of the cluster boundary is of the form:
\begin{equation}
{\mathcal H}_0^{\prime} = -\left(S_1+S_2 \right) \lambda_a -\left(S_3+S_4 \right) \lambda_b,
\label{a11}
\end{equation}
where the molecular field parameters which minimize the Gibbs energy are given by:
\begin{eqnarray}
\lambda_a&=&J_1 \left( m_a +m_b \right)+3J_2m_b \nonumber \\
\lambda_b&=&J_1 \left( m_a +m_b \right)+3J_2m_a.
\label{a12}
\end{eqnarray}
The perturbative term in the Gibbs energy is presented as:
\begin{equation}
\frac{1}{N} \left< {\mathcal H} - {\mathcal H}_0 \right>_0 =\frac{J_1}{4} \left(m_a + m_b \right)^2 + \frac{3}{2} J_2 m_a m_b.
\label{a13}
\end{equation} 
The boundary magnetizations are calculated self-consistently with the cluster trial Hamiltonian ${\mathcal H}_0^c$ from the necessary minimum conditions for the Gibbs energy, namely:
\begin{equation}
\frac{\partial G}{\partial m_{\alpha}}=0,
\label{a13a}
\end{equation}
where $\alpha = a,b$, and $G$ is given by Eq. (\ref{eq5}). From Eqs. (\ref{a13a}) we obtain:

\begin{eqnarray}
m_a&=&\frac{1}{2} \left< S_1+S_2 \right>_0 \nonumber \\
m_b&=&\frac{1}{2} \left< S_3+S_4 \right>_0.
\label{a14}
\end{eqnarray}

\section{\label{sec:appendixC}$4 \times 4$ cluster}

\begin{figure}[h]
\begin{center}
\includegraphics[width=0.95\textwidth]{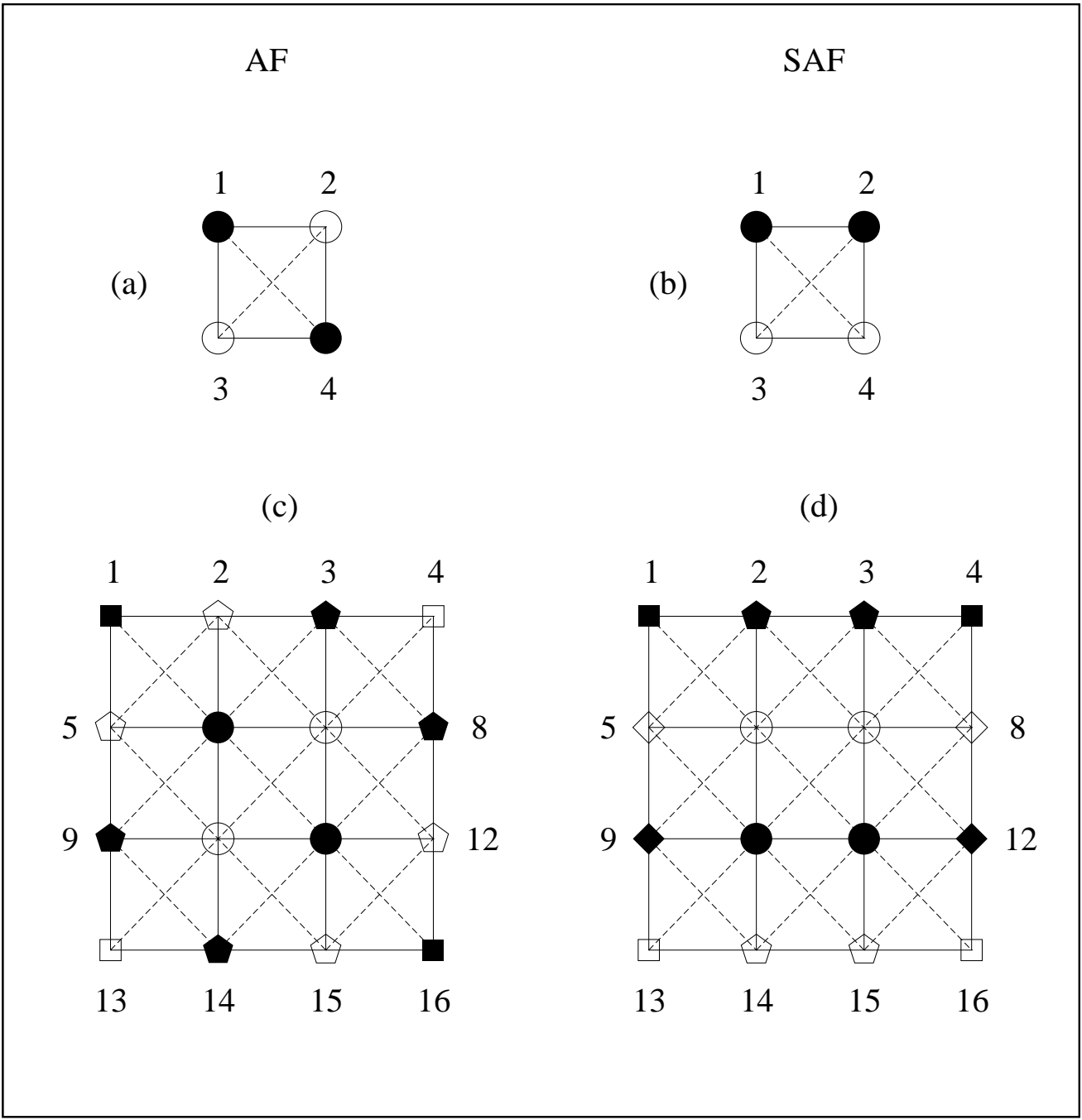}
\vspace{2mm}
\caption{\label{F1} Schematic illustration of $2 \times 2$ ((a) and (b)) as well as $4 \times 4$ ((c) and (d)) clusters. Two sublattices are marked by filled and empty symbols. AF phase is presented in figures (a) and (c), whereas SAF phase is presented in figures (b) and (d). The positions of edge spins are numbered in accordance with the corresponding formulas from Appendices \ref{sec:appendixB} and \ref{sec:appendixC}.}
\end{center}
\end{figure}

\begin{figure}[h]
\begin{center}
\includegraphics[width=0.9\textwidth]{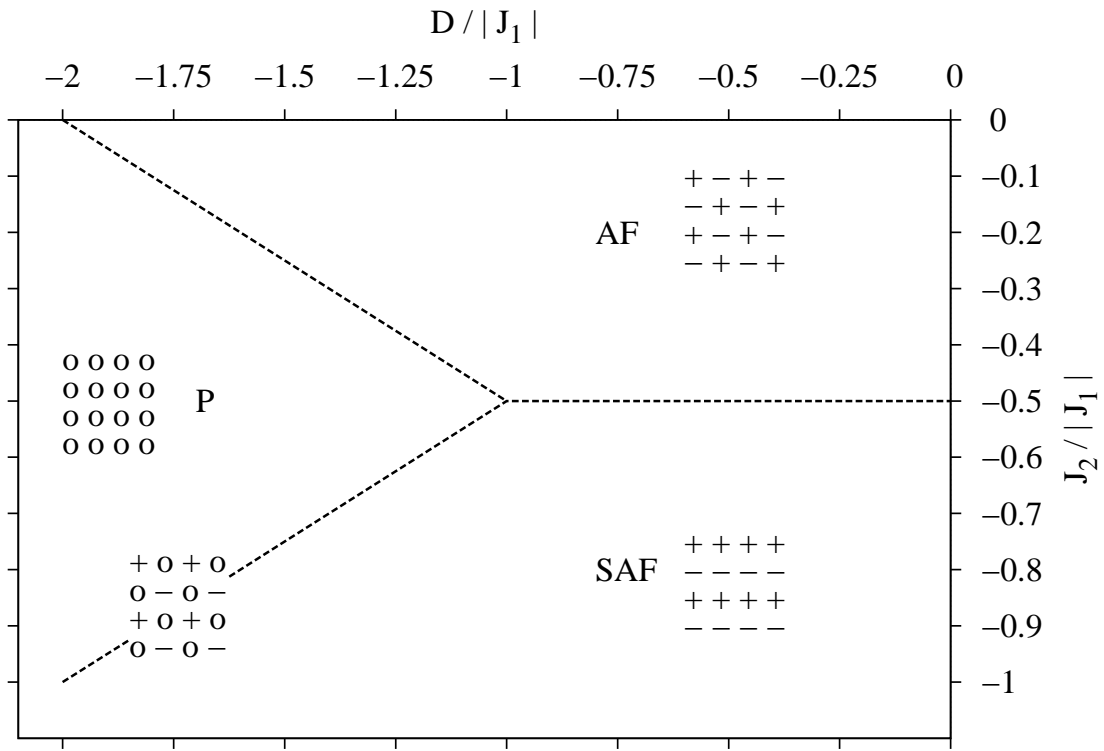}
\vspace{2mm}
\caption{\label{F2} The ground-state phase diagram, illustrating distribution of AF, SAF and P phases in the  $D/|J_1|$ - $J_2/|J_1|$ space, at $T=0$. The spin states are schematically depicted. A new mixed phase existing on P/SAF boundary is also shown.}
\end{center}
\end{figure}

\begin{figure}[h]
\begin{center}
\includegraphics[width=0.9\textwidth]{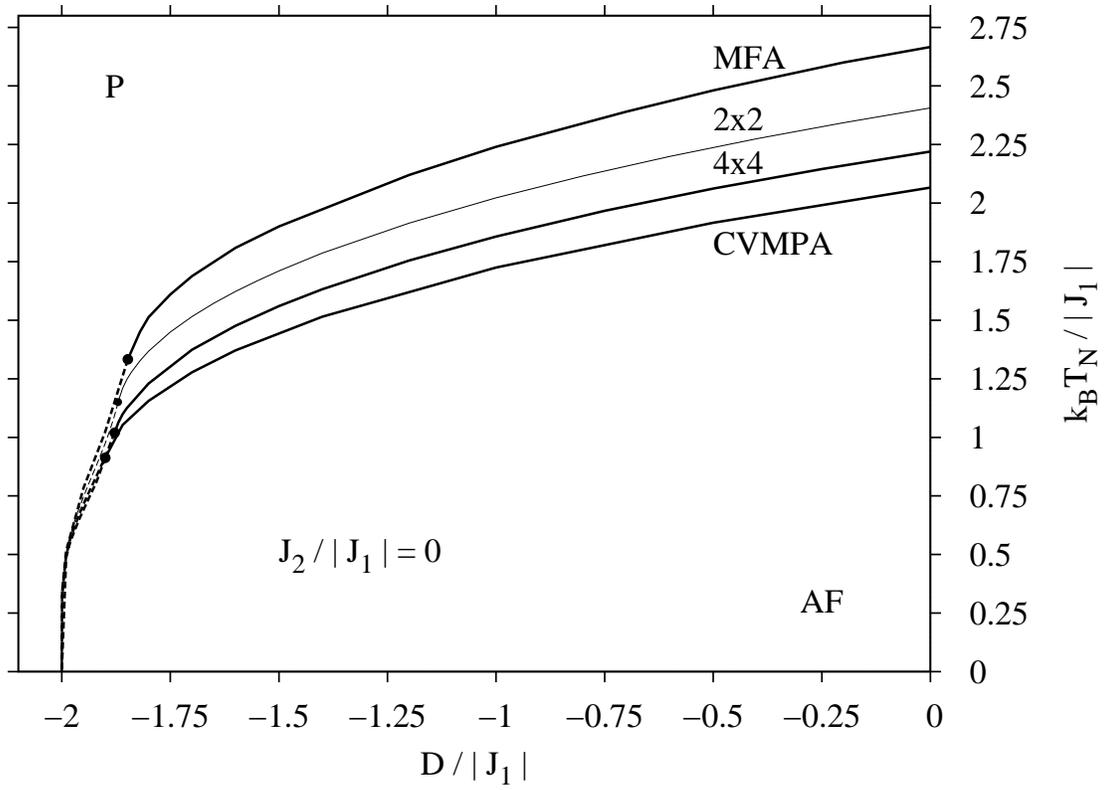}
\vspace{2mm}
\caption{\label{F3} The phase diagram showing the N\'{e}el temperatures of AF phase, $k_{\rm B}T_{\rm N}/|J_1|$, vs. single-ion anisotropy, $D/|J_1|$, for $J_2=0$, i.e., for NN interactions only (without frustrations). Results of different approximations are presented. The solid lines illustrate the continuous (2nd order) phase transitions, whereas the dashed lines are for the discontinuous (1st order) ones. The TCPs are denoted by the bold dots.}
\end{center}
\end{figure}

\begin{figure}[h]
\begin{center}
\includegraphics[width=0.9\textwidth]{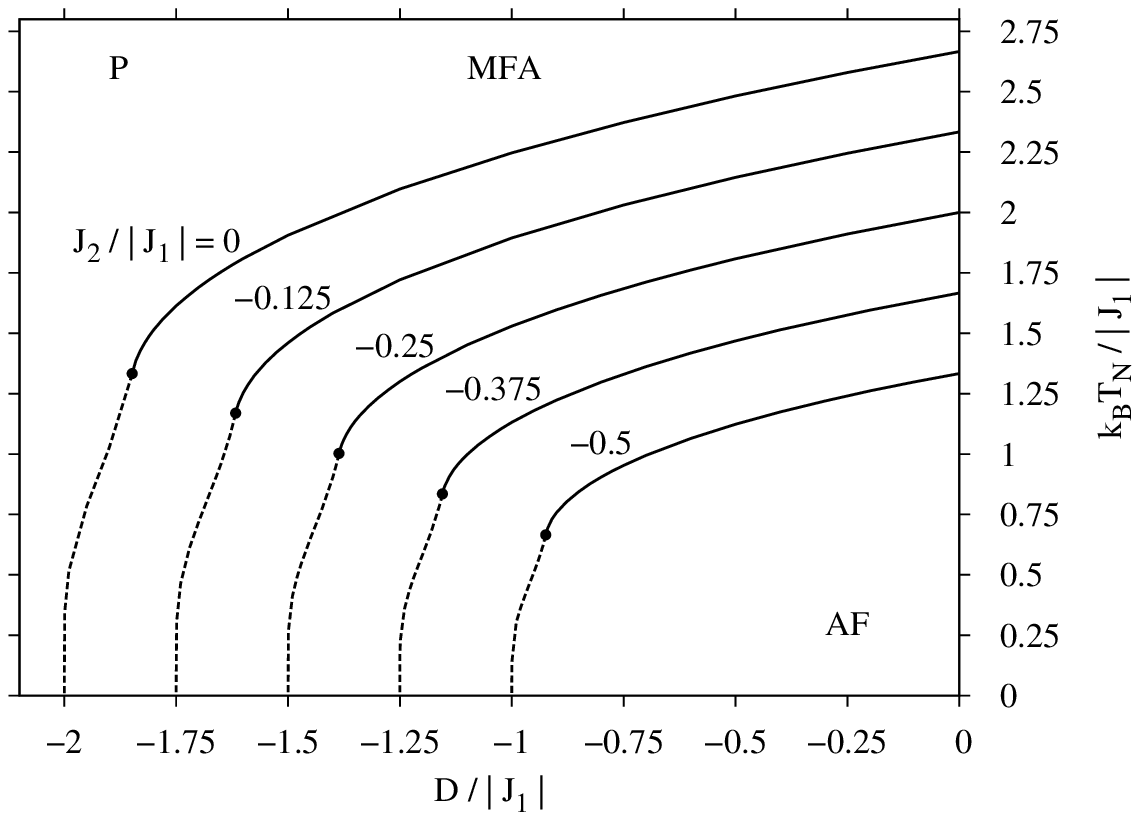}
\vspace{2mm}
\caption{\label{F4} The phase diagram in MFA showing the N\'{e}el temperatures of AF phase, $k_{\rm B}T_{\rm N}/|J_1|$, vs. single-ion anisotropy, $D/|J_1|$, for different NNN interactions $J_2/|J_1|$. Solid and dashed lines correspond to 2nd and 1st order phase transitions, respectively. TCPs positions are marked by the bold dots.}
\end{center}
\end{figure}

\begin{figure}[h]
\begin{center}
\includegraphics[width=0.9\textwidth]{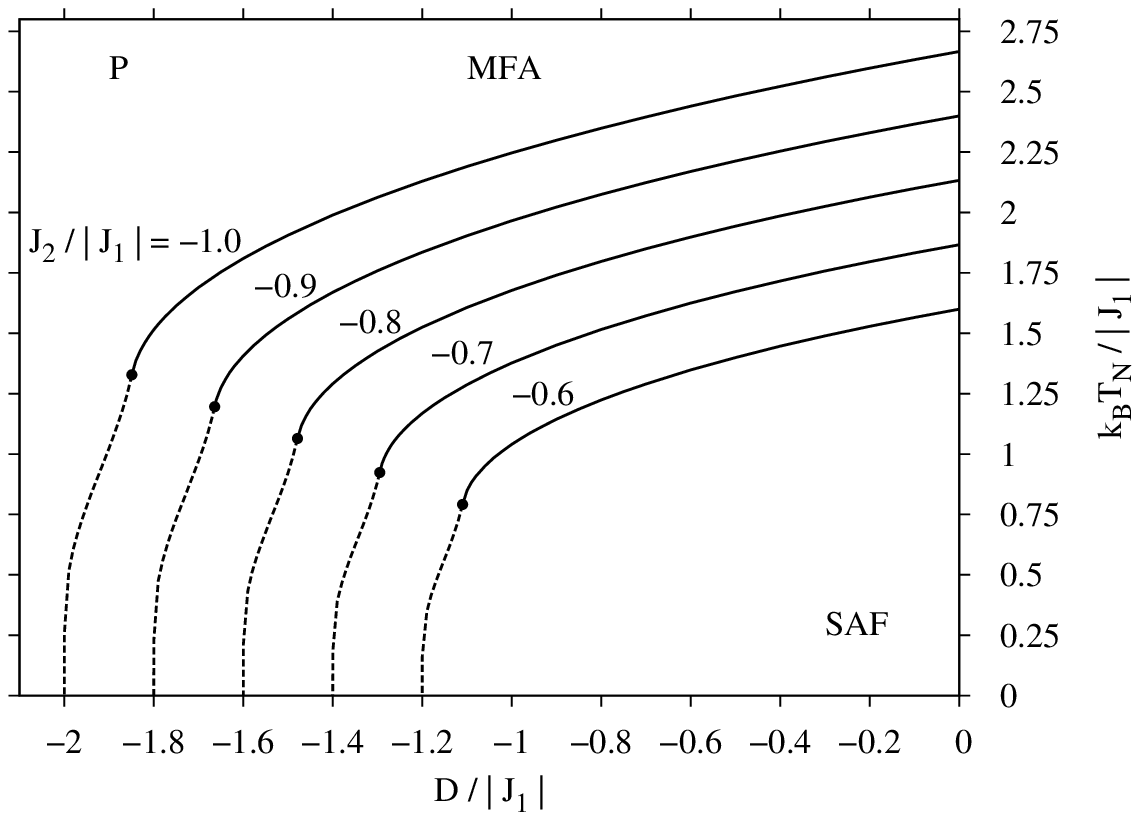}
\vspace{2mm}
\caption{\label{F5} The phase diagram in MFA showing the N\'{e}el temperatures of SAF phase, $k_{\rm B}T_{\rm N}/|J_1|$, vs. single-ion anisotropy, $D/|J_1|$, for different NNN interactions $J_2/|J_1|$. Solid and dashed lines correspond to 2nd and 1st order phase transitions, respectively. TCPs positions are marked by the bold dots.}
\end{center}
\end{figure}

\begin{figure}[h]
\begin{center}
\includegraphics[width=0.9\textwidth]{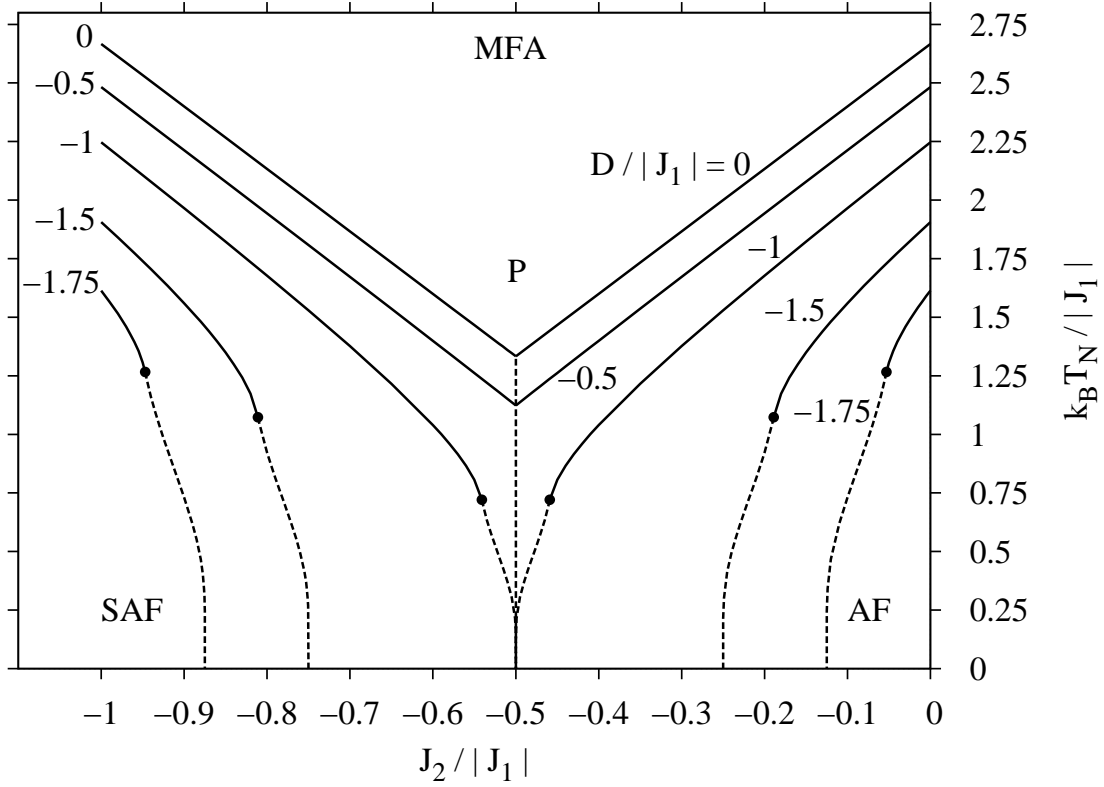}
\vspace{2mm}
\caption{\label{F6} The phase diagram in MFA showing the N\'{e}el temperatures of AF and SAF phases, $k_{\rm B}T_{\rm N}/|J_1|$, vs. NNN interaction parameter, $J_2/|J_1|$, for different single-ion anisotropies, $D/|J_1|$. Solid and dashed lines correspond to 2nd and 1st order phase transitions, respectively. TCPs positions are marked by the bold dots. At $J_2/|J_1|=-0.5$, for the curves with $D/|J_1|=0$ and $D/|J_1|=-0.5$, the triple points exist for $T>0$, connecting the AF, SAF and P phases in equilibrium.} 
\end{center}
\end{figure}

\begin{figure}[h]
\begin{center}
\includegraphics[width=0.9\textwidth]{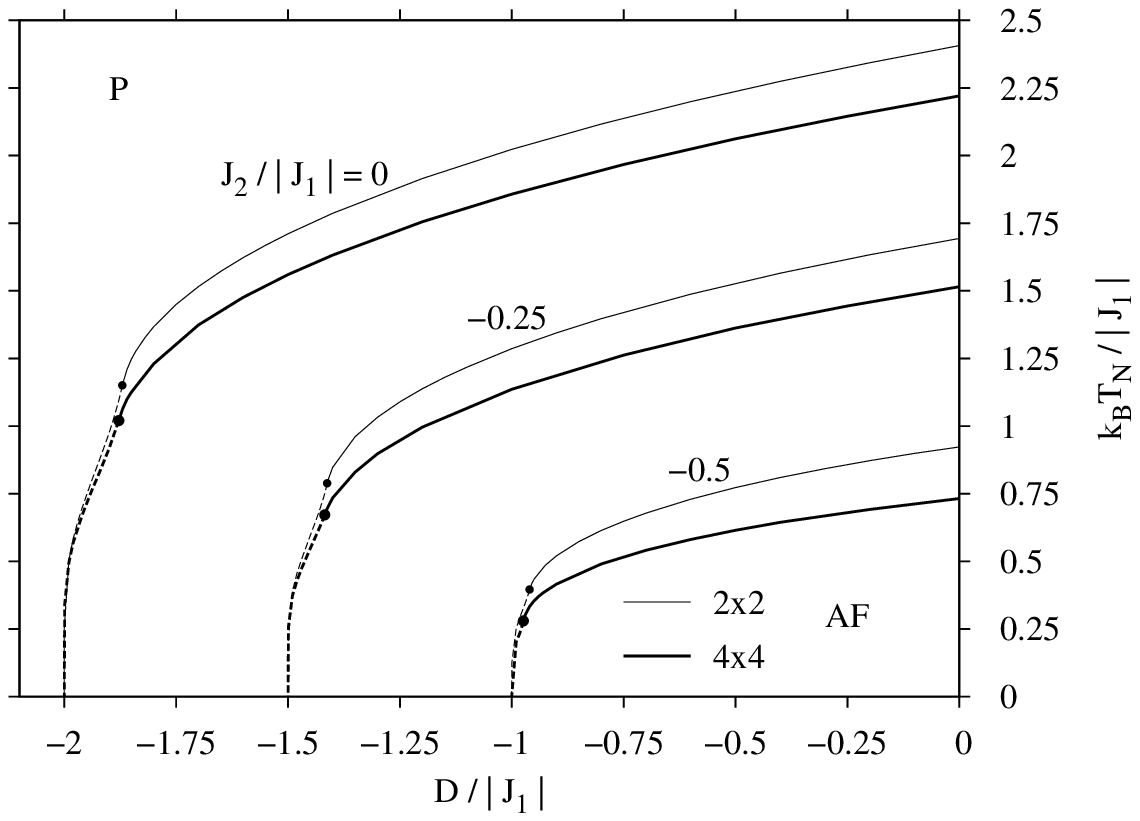}
\vspace{2mm}
\caption{\label{F7} The phase diagram showing the N\'{e}el temperatures of AF phase, $k_{\rm B}T_{\rm N}/|J_1|$, vs. single-ion anisotropy, $D/|J_1|$, for different NNN interactions $J_2/|J_1|$.
Thick lines correspond to the approximation based on $4 \times 4$ clusters, whereas thin lines are for $2 \times 2$ clusters.
Solid and dashed lines correspond to 2nd and 1st order phase transitions, respectively. TCPs positions are marked by the bold dots.}
\end{center}
\end{figure}

\begin{figure}[h]
\begin{center}
\includegraphics[width=0.9\textwidth]{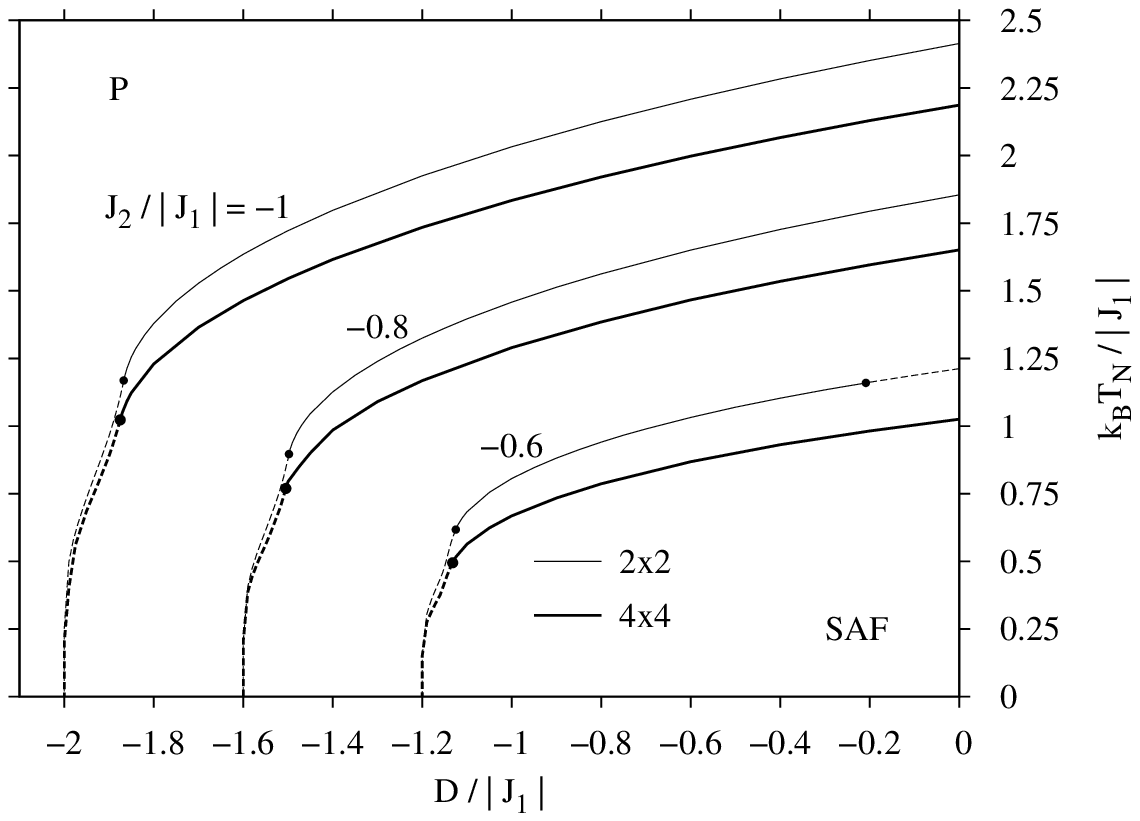}
\vspace{2mm}
\caption{\label{F8} The phase diagram showing the N\'{e}el temperatures of SAF phase, $k_{\rm B}T_{\rm N}/|J_1|$, vs. single-ion anisotropy, $D/|J_1|$, for different NNN interactions $J_2/|J_1|$.
Thick lines correspond to the approximation based on $4 \times 4$ clusters, whereas thin lines are for $2 \times 2$ clusters.
Solid and dashed lines correspond to 2nd and 1st order phase transitions, respectively. TCPs positions are marked by the bold dots.}
\end{center}
\end{figure}

\begin{figure}[h]
\begin{center}
\includegraphics[width=0.9\textwidth]{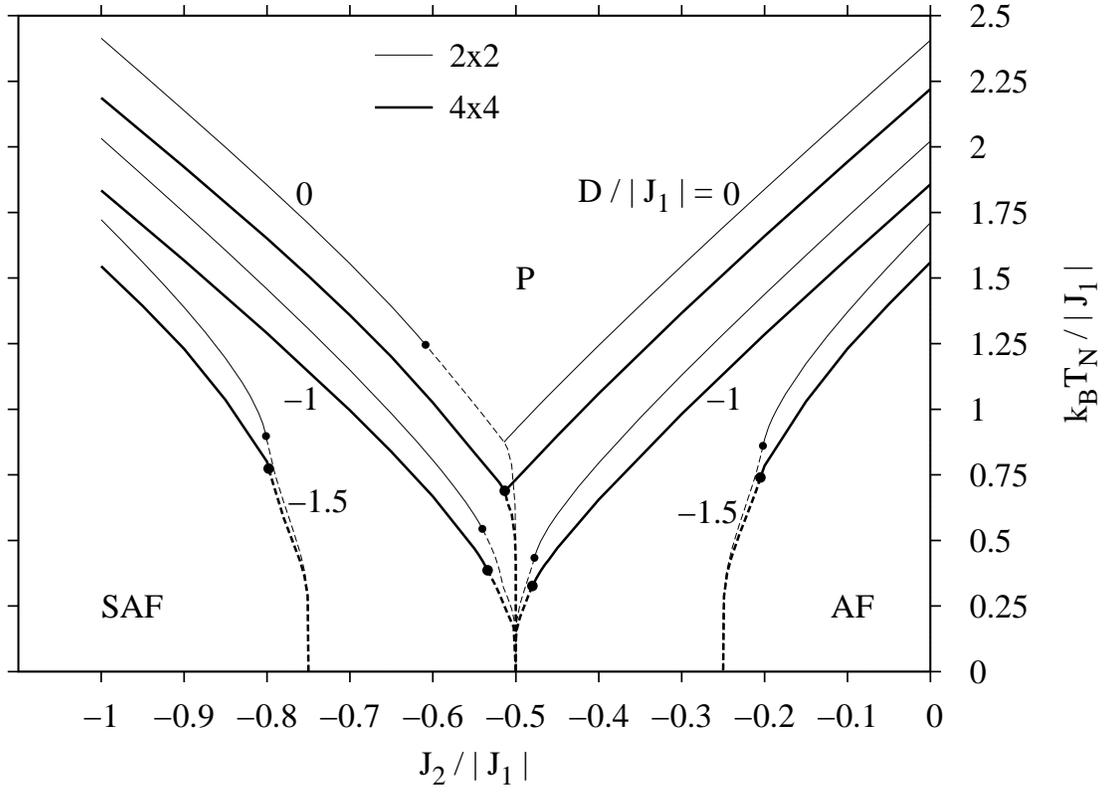}
\vspace{2mm}
\caption{\label{F9} The phase diagram showing the N\'{e}el temperatures of AF and SAF phases, $k_{\rm B}T_{\rm N}/|J_1|$, vs. NNN interaction parameter, $J_2/|J_1|$, for different single-ion anisotropies, $D/|J_1|$. 
Thick lines correspond to the approximation based on $4 \times 4$ clusters, whereas thin lines are for $2 \times 2$ clusters.
Solid and dashed lines correspond to 2nd and 1st order phase transitions, respectively. TCPs positions are marked by the bold dots. The triple points existing for $T>0$, for the curves labelled by $D/|J_1|=0$, are shifted towards $J_2/|J_1|<-0.5$.}
\end{center}
\end{figure}

\begin{figure}[h]
\begin{center}
\includegraphics[width=0.9\textwidth]{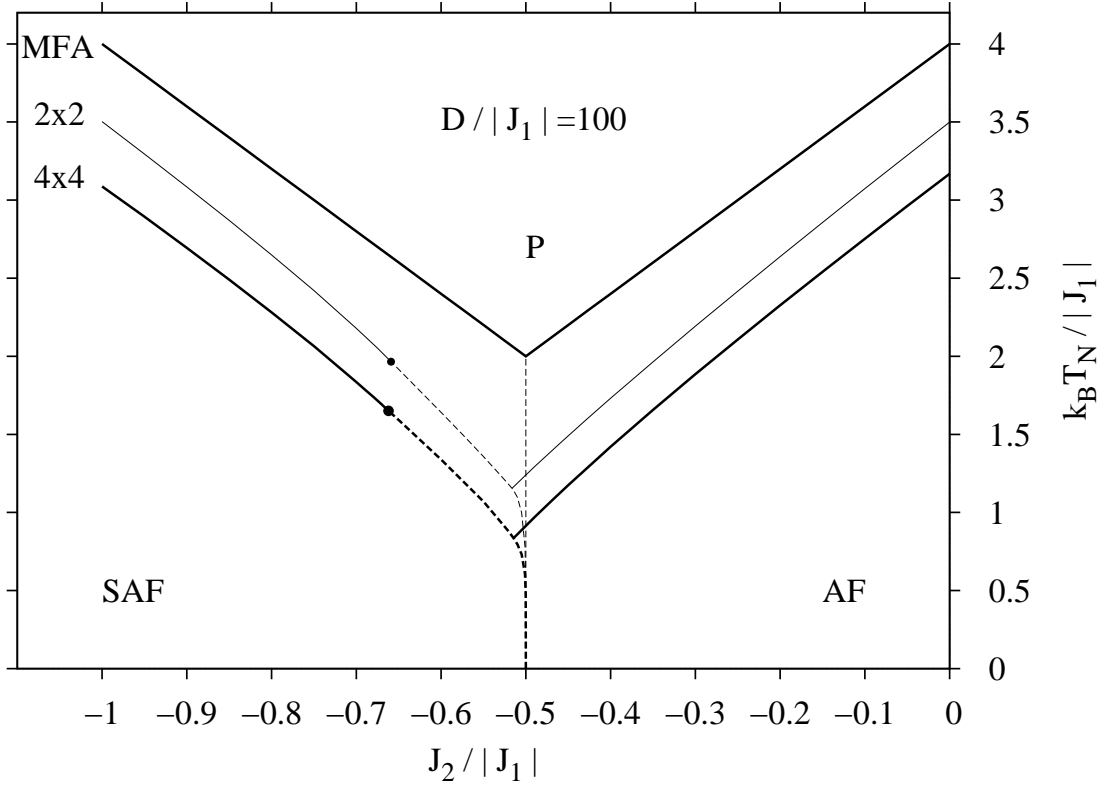}
\vspace{2mm}
\caption{\label{F10} The phase diagram showing the N\'{e}el temperatures of AF and SAF phases, $k_{\rm B}T_{\rm N}/|J_1|$, vs. NNN interaction parameter, $J_2/|J_1|$, for strong, and positive, single-ion anisotropy, $D/|J_1|=100$. 
Thick lines correspond to the approximation based on MFA and $4 \times 4$ cluster, whereas thin lines are for $2 \times 2$ cluster.
Solid and dashed lines correspond to 2nd and 1st order phase transitions, respectively. TCPs positions are marked by the bold dots. In case of the approximations based on $4 \times 4$ and $2 \times 2$ clusters, the triple points are shifted towards $J_2/|J_1|<-0.5$.}
\end{center}
\end{figure}

\begin{figure}[h]
\begin{center}
\includegraphics[width=0.9\textwidth]{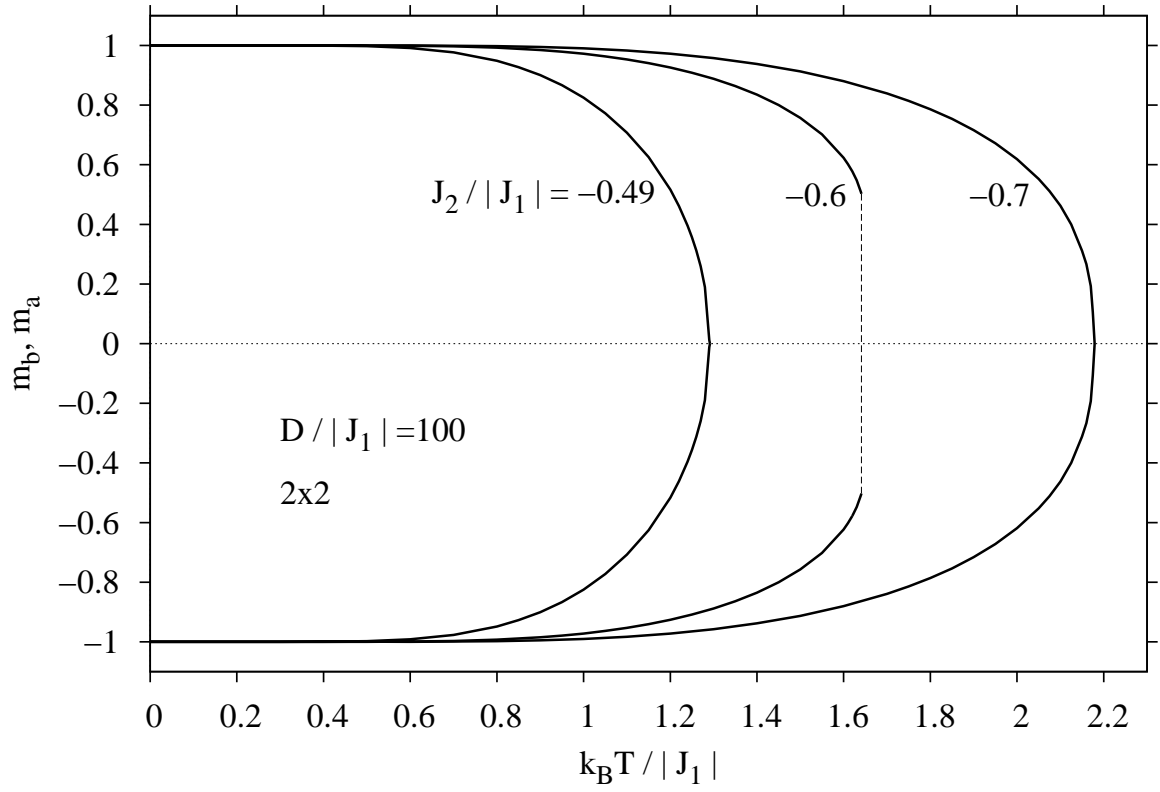}
\vspace{2mm}
\caption{\label{F11} The sublattice magnetizations $m_a$ and $m_b$ vs. dimensionless temperature $k_{\rm B}T/|J_1|$ for $2 \times 2$ cluster with single-ion anisotropy $D/|J_1|=100$. The curves are plotted for $J_2/|J_1|=-0.49$, -0.6 and -0.7.}
\end{center}
\end{figure}

\underline
{AF - phase:} (see Fig.~\ref{F1} (c))\\

Hamiltonian of the cluster boundary is of the form:
\begin{eqnarray}
{\mathcal H}_0^{\prime} = &-&\left(S_1+S_{16} \right) \lambda_a -\left(S_4+S_{13} \right) \lambda_b
\nonumber \\
&-&\left(S_3+S_8+S_9+S_{14} \right) \nu_a -\left(S_2+S_5+S_{12}+S_{15} \right) \nu_b,
\label{a15}
\end{eqnarray}
where the molecular field parameters which minimize the Gibbs energy are given by:
\begin{eqnarray}
\lambda_a&=&2J_1m_b^c+2J_2m_a^s+J_2m_a^c \nonumber \\
\lambda_b&=&2J_1m_a^c+2J_2m_b^s+J_2m_b^c \nonumber \\
\nu_a&=&J_1m_b^s+J_2 \left(m_a^c+m_a^s\right) \nonumber \\
\nu_b&=&J_1m_a^s+J_2 \left(m_b^c+m_b^s\right).
\label{a16}
\end{eqnarray}
The perturbative term in the Gibbs energy is presented as:
\begin{eqnarray}
\frac{1}{N} \left< {\mathcal H} - {\mathcal H}_0 \right>_0 &=&\frac{1}{4}J_1 \left(m_a^c m_b^c+m_a^s m_b^s\right) + \frac{1}{4}J_2 \left(m_a^c m_a^s+m_b^c m_b^s\right)\nonumber \\
 &+& \frac{1}{16} J_2 \left( 2\left( m_a^s\right)^2 + 2\left( m_b^s\right)^2 +\left( m_a^c\right)^2 + \left( m_b^c\right)^2 \right).
\label{a17}
\end{eqnarray} 
The boundary magnetizations are calculated self-consistently with the cluster trial Hamiltonian ${\mathcal H}_0^c$ from the necessary minimum conditions for the Gibbs energy, namely:
\begin{equation}
\frac{\partial G}{\partial m_{\alpha}^x}=0,
\label{a17a}
\end{equation}
where $\alpha = a,b$, $x=c,s$ and $G$ is given by Eq. (\ref{eq5}). Index $x=c,s$ corresponds to two non-equivalent edge spin positions in Fig.~\ref{F1}(c): $c$ - corner spins and $s$ - side spins. 
From Eqs. (\ref{a17a}) we obtain:

\begin{eqnarray}
m_a^c&=&\frac{1}{2} \left< S_1+S_{16} \right>_0 \nonumber \\
m_b^c&=&\frac{1}{2} \left< S_4+S_{13} \right>_0 \nonumber \\
m_a^s&=&\frac{1}{4} \left< S_3+S_8+S_9+S_{14} \right>_0 \nonumber \\
m_b^s&=&\frac{1}{4} \left< S_2+S_5+S_{12}+S_{15} \right>_0.
\label{a18}
\end{eqnarray}

\underline
{SAF - phase:} (see Fig.~\ref{F1} (d))\\

Hamiltonian of the cluster boundary is of the form:
\begin{eqnarray}
{\mathcal H}_0^{\prime} = &-&\left(S_1+S_{4} \right) \lambda_a -\left(S_{13}+S_{16} \right)
\lambda_b
\nonumber \\
&-&\left(S_2+S_3 \right) \nu_a -\left(S_{14}+S_{15} \right) \nu_b
\nonumber \\
&-&\left(S_9+S_{12} \right) \mu_a -\left(S_5+S_8 \right) \mu_b,
\label{a19}
\end{eqnarray}
where the molecular field parameters which minimize the Gibbs energy are given by:
\begin{eqnarray}
\lambda_a&=&J_1\left(m_a^c+m_b^c\right)+J_2\left(m_b^c+m_b^h+m_b^v \right) \nonumber \\
\lambda_b&=&J_1\left(m_a^c+m_b^c\right)+J_2\left(m_a^c+m_a^h+m_a^v \right) \nonumber \\
\nu_a&=&J_1m_b^h+J_2 \left(m_b^c+m_b^h\right) \nonumber \\
\nu_b&=&J_1m_a^h+J_2 \left(m_a^c+m_a^h\right) \nonumber \\
\mu_a&=&J_1m_a^v+J_2 \left(m_b^c+m_b^v\right) \nonumber \\
\mu_b&=&J_1m_b^v+J_2 \left(m_a^c+m_a^v\right).
\label{a20}
\end{eqnarray}
The perturbative term in the Gibbs energy is presented as:
\begin{eqnarray}
\frac{1}{N} \left< {\mathcal H} - {\mathcal H}_0 \right>_0 &=&\frac{1}{16}J_1 \left(2m_a^c m_b^c+2m_a^h m_b^h +\left( m_a^c\right)^2 +\left( m_b^c\right)^2 +\left( m_a^v\right)^2 +\left( m_b^v\right)^2\right) \nonumber \\
 &+& \frac{1}{8} J_2 \left( m_a^c m_b^c+ m_a^c m_b^h+ m_a^h m_b^c + m_a^h m_b^h+ m_a^v m_b^v+ m_a^c m_b^v+ m_a^v m_b^c \right). \nonumber\\
\label{a21}
\end{eqnarray} 
The boundary magnetizations are calculated self-consistently with the cluster trial Hamiltonian ${\mathcal H}_0^c$ from the necessary minimum conditions for the Gibbs energy, namely:
\begin{equation}
\frac{\partial G}{\partial m_{\alpha}^y}=0,
\label{a21a}
\end{equation}
where $\alpha = a,b$, $y=c,h,v$ and $G$ is given by Eq. (\ref{eq5}). Index $y=c,h,v$ corresponds to three non-equivalent edge spin positions in Fig.~\ref{F1}(d): $c$ - corner spins and two types of side spins: $h$ - on the horizontal sides and $v$ - on the vertical sides. 
From Eqs. (\ref{a21a}) we obtain:

\begin{eqnarray}
m_a^c&=&\frac{1}{2} \left< S_1+S_4 \right>_0 \nonumber \\
m_b^c&=&\frac{1}{2} \left< S_{13}+S_{16} \right>_0 \nonumber \\
m_a^h&=&\frac{1}{2} \left< S_2+S_3 \right>_0 \nonumber \\
m_b^h&=&\frac{1}{2} \left< S_{14}+S_{15} \right>_0 \nonumber \\
m_a^v&=&\frac{1}{2} \left< S_9+S_{12} \right>_0 \nonumber \\
m_b^v&=&\frac{1}{2} \left< S_5+S_8 \right>_0.
\label{a22}
\end{eqnarray}

%


\begin{thebibliography}{41}%
\makeatletter
\providecommand \@ifxundefined [1]{%
 \@ifx{#1\undefined}
}%
\providecommand \@ifnum [1]{%
 \ifnum #1\expandafter \@firstoftwo
 \else \expandafter \@secondoftwo
 \fi
}%
\providecommand \@ifx [1]{%
 \ifx #1\expandafter \@firstoftwo
 \else \expandafter \@secondoftwo
 \fi
}%
\providecommand \natexlab [1]{#1}%
\providecommand \enquote  [1]{``#1''}%
\providecommand \bibnamefont  [1]{#1}%
\providecommand \bibfnamefont [1]{#1}%
\providecommand \citenamefont [1]{#1}%
\providecommand \href@noop [0]{\@secondoftwo}%
\providecommand \href [0]{\begingroup \@sanitize@url \@href}%
\providecommand \@href[1]{\@@startlink{#1}\@@href}%
\providecommand \@@href[1]{\endgroup#1\@@endlink}%
\providecommand \@sanitize@url [0]{\catcode `\\12\catcode `\$12\catcode
  `\&12\catcode `\#12\catcode `\^12\catcode `\_12\catcode `\%12\relax}%
\providecommand \@@startlink[1]{}%
\providecommand \@@endlink[0]{}%
\providecommand \url  [0]{\begingroup\@sanitize@url \@url }%
\providecommand \@url [1]{\endgroup\@href {#1}{\urlprefix }}%
\providecommand \urlprefix  [0]{URL }%
\providecommand \Eprint [0]{\href }%
\providecommand \doibase [0]{http://dx.doi.org/}%
\providecommand \selectlanguage [0]{\@gobble}%
\providecommand \bibinfo  [0]{\@secondoftwo}%
\providecommand \bibfield  [0]{\@secondoftwo}%
\providecommand \translation [1]{[#1]}%
\providecommand \BibitemOpen [0]{}%
\providecommand \bibitemStop [0]{}%
\providecommand \bibitemNoStop [0]{.\EOS\space}%
\providecommand \EOS [0]{\spacefactor3000\relax}%
\providecommand \BibitemShut  [1]{\csname bibitem#1\endcsname}%
\let\auto@bib@innerbib\@empty
\bibitem [{\citenamefont {Binder}\ and\ \citenamefont
  {Landau}(1980)}]{Binder1980}%
  \BibitemOpen
  \bibfield  {author} {\bibinfo {author} {\bibfnamefont {K.}~\bibnamefont
  {Binder}}\ and\ \bibinfo {author} {\bibfnamefont {D.~P.}\ \bibnamefont
  {Landau}},\ }\bibfield  {title} {\enquote {\bibinfo {title} {Phase
  {{Diagrams}} and {{Critical Behavior}} in {{Ising Square Lattices}} with
  {{Nearest}}- and next-{{Nearest}}-{{Neighbor Interactions}}},}\ }\href
  {\doibase 10.1103/PhysRevB.21.1941} {\bibfield  {journal} {\bibinfo
  {journal} {Physical Review B}\ }\textbf {\bibinfo {volume} {21}},\ \bibinfo
  {pages} {1941--1962} (\bibinfo {year} {1980})}\BibitemShut {NoStop}%
\bibitem [{\citenamefont {Landau}(1980)}]{Landau1980}%
  \BibitemOpen
  \bibfield  {author} {\bibinfo {author} {\bibfnamefont {D.~P.}\ \bibnamefont
  {Landau}},\ }\bibfield  {title} {\enquote {\bibinfo {title} {Phase
  {{Transitions}} in the {{Ising Square Lattice}} with
  {{Next}}-{{Nearest}}-{{Neighbor Interactions}}},}\ }\href {\doibase
  10.1103/PhysRevB.21.1285} {\bibfield  {journal} {\bibinfo  {journal}
  {Physical Review B}\ }\textbf {\bibinfo {volume} {21}},\ \bibinfo {pages}
  {1285--1297} (\bibinfo {year} {1980})}\BibitemShut {NoStop}%
\bibitem [{\citenamefont {Jin}\ \emph {et~al.}(2013)\citenamefont {Jin},
  \citenamefont {Sen}, \citenamefont {Guo},\ and\ \citenamefont
  {Sandvik}}]{Jin2013}%
  \BibitemOpen
  \bibfield  {author} {\bibinfo {author} {\bibfnamefont {Songbo}\ \bibnamefont
  {Jin}}, \bibinfo {author} {\bibfnamefont {Arnab}\ \bibnamefont {Sen}},
  \bibinfo {author} {\bibfnamefont {Wenan}\ \bibnamefont {Guo}}, \ and\
  \bibinfo {author} {\bibfnamefont {Anders~W.}\ \bibnamefont {Sandvik}},\
  }\bibfield  {title} {\enquote {\bibinfo {title} {Phase {{Transitions}} in the
  {{Frustrated Ising Model}} on the {{Square Lattice}}},}\ }\href {\doibase
  10.1103/PhysRevB.87.144406} {\bibfield  {journal} {\bibinfo  {journal}
  {Physical Review B}\ }\textbf {\bibinfo {volume} {87}},\ \bibinfo {pages}
  {144406} (\bibinfo {year} {2013})}\BibitemShut {NoStop}%
\bibitem [{\citenamefont {Samuelsen}(1974)}]{Samuelsen1974}%
  \BibitemOpen
  \bibfield  {author} {\bibinfo {author} {\bibfnamefont {E.~J.}\ \bibnamefont
  {Samuelsen}},\ }\bibfield  {title} {\enquote {\bibinfo {title} {Critical
  {{Behaviour}} of the {{Two}}-{{Dimensional Ising Antiferromagnets K$_2$CoF$_4$}}
  and {{Rb$_2$CoF$_4$}}},}\ }\href {\doibase 10.1016/S0022-3697(74)80258-1}
  {\bibfield  {journal} {\bibinfo  {journal} {Journal of Physics and Chemistry
  of Solids}\ }\textbf {\bibinfo {volume} {35}},\ \bibinfo {pages} {785--793}
  (\bibinfo {year} {1974})}\BibitemShut {NoStop}%
\bibitem [{\citenamefont {Ikeda}\ and\ \citenamefont
  {Hirakawa}(1974)}]{Ikeda1974}%
  \BibitemOpen
  \bibfield  {author} {\bibinfo {author} {\bibfnamefont {Hironobu}\
  \bibnamefont {Ikeda}}\ and\ \bibinfo {author} {\bibfnamefont {Kinshiro}\
  \bibnamefont {Hirakawa}},\ }\bibfield  {title} {\enquote {\bibinfo {title}
  {Neutron {{Scattering Study}} of {{Two}}-{{Dimensional Ising Nature}} of
  {{K$_2$CoF$_4$}}},}\ }\href {\doibase 10.1016/0038-1098(74)91004-7} {\bibfield
  {journal} {\bibinfo  {journal} {Solid State Communications}\ }\textbf
  {\bibinfo {volume} {14}},\ \bibinfo {pages} {529--532} (\bibinfo {year}
  {1974})}\BibitemShut {NoStop}%
\bibitem [{\citenamefont {{de Jongh}}\ and\ \citenamefont
  {Miedema}(1974)}]{Jongh1974}%
  \BibitemOpen
  \bibfield  {author} {\bibinfo {author} {\bibfnamefont {L.~J.}\ \bibnamefont
  {{de Jongh}}}\ and\ \bibinfo {author} {\bibfnamefont {A.~R.}\ \bibnamefont
  {Miedema}},\ }\bibfield  {title} {\enquote {\bibinfo {title} {Experiments on
  {{Simple Magnetic Model Systems}}},}\ }\href {\doibase
  10.1080/00018739700101558} {\bibfield  {journal} {\bibinfo  {journal}
  {Advances in Physics}\ }\textbf {\bibinfo {volume} {23}},\ \bibinfo {pages}
  {1--260} (\bibinfo {year} {1974})}\BibitemShut {NoStop}%
\bibitem [{\citenamefont {Mor{\'a}n-L{\'o}pez}\ \emph
  {et~al.}(1993)\citenamefont {Mor{\'a}n-L{\'o}pez}, \citenamefont
  {Aguilera-Granja},\ and\ \citenamefont {Sanchez}}]{Moran-Lopez1993}%
  \BibitemOpen
  \bibfield  {author} {\bibinfo {author} {\bibfnamefont {J.~L.}\ \bibnamefont
  {Mor{\'a}n-L{\'o}pez}}, \bibinfo {author} {\bibfnamefont {F.}~\bibnamefont
  {Aguilera-Granja}}, \ and\ \bibinfo {author} {\bibfnamefont {J.~M.}\
  \bibnamefont {Sanchez}},\ }\bibfield  {title} {\enquote {\bibinfo {title}
  {First-{{Order Phase Transitions}} in the {{Ising Square Lattice}} with
  {{First}}- and {{Second}}-{{Neighbor Interactions}}},}\ }\href {\doibase
  10.1103/PhysRevB.48.3519} {\bibfield  {journal} {\bibinfo  {journal}
  {Physical Review B}\ }\textbf {\bibinfo {volume} {48}},\ \bibinfo {pages}
  {3519--3522} (\bibinfo {year} {1993})}\BibitemShut {NoStop}%
\bibitem [{\citenamefont {Bob{\'a}k}\ \emph {et~al.}(1996)\citenamefont
  {Bob{\'a}k}, \citenamefont {Mockov{\v c}iak}, \citenamefont {Jur{\v c}i{\v
  s}in},\ and\ \citenamefont {Ja{\v s}{\v c}ur}}]{Bobak1996}%
  \BibitemOpen
  \bibfield  {author} {\bibinfo {author} {\bibfnamefont {A.}~\bibnamefont
  {Bob{\'a}k}}, \bibinfo {author} {\bibfnamefont {S.}~\bibnamefont {Mockov{\v
  c}iak}}, \bibinfo {author} {\bibfnamefont {M.}~\bibnamefont {Jur{\v c}i{\v
  s}in}}, \ and\ \bibinfo {author} {\bibfnamefont {M.}~\bibnamefont {Ja{\v
  s}{\v c}ur}},\ }\bibfield  {title} {\enquote {\bibinfo {title} {Tricritical
  {{Behaviour}} of the {{Ferromagnetic Blume}}-{{Capel Model}} with {{First}}-
  and {{Second}}-{{Neighbour Interactions}}},}\ }\href {\doibase
  10.1016/0378-4371(96)00067-2} {\bibfield  {journal} {\bibinfo  {journal}
  {Physica A: Statistical Mechanics and its Applications}\ }\textbf {\bibinfo
  {volume} {230}},\ \bibinfo {pages} {703--712} (\bibinfo {year}
  {1996})}\BibitemShut {NoStop}%
\bibitem [{\citenamefont {{dos Anjos}}\ \emph {et~al.}(2008)\citenamefont {{dos
  Anjos}}, \citenamefont {Roberto~Viana},\ and\ \citenamefont {{Ricardo de
  Sousa}}}]{dosAnjos2008}%
  \BibitemOpen
  \bibfield  {author} {\bibinfo {author} {\bibfnamefont {Rosana~A.}\
  \bibnamefont {{dos Anjos}}}, \bibinfo {author} {\bibfnamefont
  {J.}~\bibnamefont {Roberto~Viana}}, \ and\ \bibinfo {author} {\bibfnamefont
  {J.}~\bibnamefont {{Ricardo de Sousa}}},\ }\bibfield  {title} {\enquote
  {\bibinfo {title} {Phase {{Diagram}} of the {{Ising Antiferromagnet}} with
  {{Nearest}}-{{Neighbor}} and next-{{Nearest}}-{{Neighbor Interactions}} on a
  {{Square Lattice}}},}\ }\href {\doibase 10.1016/j.physleta.2007.09.059}
  {\bibfield  {journal} {\bibinfo  {journal} {Physics Letters A}\ }\textbf
  {\bibinfo {volume} {372}},\ \bibinfo {pages} {1180--1184} (\bibinfo {year}
  {2008})}\BibitemShut {NoStop}%
\bibitem [{\citenamefont {Kalz}\ \emph {et~al.}(2008)\citenamefont {Kalz},
  \citenamefont {Honecker}, \citenamefont {Fuchs},\ and\ \citenamefont
  {Pruschke}}]{Kalz2008}%
  \BibitemOpen
  \bibfield  {author} {\bibinfo {author} {\bibfnamefont {A.}~\bibnamefont
  {Kalz}}, \bibinfo {author} {\bibfnamefont {A.}~\bibnamefont {Honecker}},
  \bibinfo {author} {\bibfnamefont {S.}~\bibnamefont {Fuchs}}, \ and\ \bibinfo
  {author} {\bibfnamefont {T.}~\bibnamefont {Pruschke}},\ }\bibfield  {title}
  {\enquote {\bibinfo {title} {Phase {{Diagram}} of the {{Ising Square
  Lattice}} with {{Competing Interactions}}},}\ }\href {\doibase
  10.1140/epjb/e2008-00359-6} {\bibfield  {journal} {\bibinfo  {journal} {The
  European Physical Journal B}\ }\textbf {\bibinfo {volume} {65}},\ \bibinfo
  {pages} {533} (\bibinfo {year} {2008})}\BibitemShut {NoStop}%
\bibitem [{\citenamefont {Bob{\'a}k}\ \emph {et~al.}(2015)\citenamefont
  {Bob{\'a}k}, \citenamefont {Lu{\v c}ivjansk{\'y}}, \citenamefont
  {Borovsk{\'y}},\ and\ \citenamefont {{\v Z}ukovi{\v c}}}]{Bobak2015}%
  \BibitemOpen
  \bibfield  {author} {\bibinfo {author} {\bibfnamefont {A.}~\bibnamefont
  {Bob{\'a}k}}, \bibinfo {author} {\bibfnamefont {T.}~\bibnamefont {Lu{\v
  c}ivjansk{\'y}}}, \bibinfo {author} {\bibfnamefont {M.}~\bibnamefont
  {Borovsk{\'y}}}, \ and\ \bibinfo {author} {\bibfnamefont {M.}~\bibnamefont
  {{\v Z}ukovi{\v c}}},\ }\bibfield  {title} {\enquote {\bibinfo {title} {Phase
  {{Transitions}} in a {{Frustrated Ising Antiferromagnet}} on a {{Square
  Lattice}}},}\ }\href {\doibase 10.1103/PhysRevE.91.032145} {\bibfield
  {journal} {\bibinfo  {journal} {Physical Review E}\ }\textbf {\bibinfo
  {volume} {91}},\ \bibinfo {pages} {032145} (\bibinfo {year}
  {2015})}\BibitemShut {NoStop}%
\bibitem [{\citenamefont {Murtazaev}\ \emph {et~al.}(2014)\citenamefont
  {Murtazaev}, \citenamefont {Ramazanov},\ and\ \citenamefont
  {Kassan-Ogly}}]{Murtazaev2014}%
  \BibitemOpen
  \bibfield  {author} {\bibinfo {author} {\bibfnamefont {A.~K.}\ \bibnamefont
  {Murtazaev}}, \bibinfo {author} {\bibfnamefont {M.~K.}\ \bibnamefont
  {Ramazanov}}, \ and\ \bibinfo {author} {\bibfnamefont {F.~A.}\ \bibnamefont
  {Kassan-Ogly}},\ }\bibfield  {title} {\enquote {\bibinfo {title}
  {Frustrations and {{Phase Transitions}} in the {{Ising Model}} on {{Square
  Lattice}}},}\ }\href {\doibase 10.1088/1742-6596/510/1/012026} {\bibfield
  {journal} {\bibinfo  {journal} {Journal of Physics: Conference Series}\
  }\textbf {\bibinfo {volume} {510}},\ \bibinfo {pages} {012026} (\bibinfo
  {year} {2014})}\BibitemShut {NoStop}%
\bibitem [{\citenamefont {Richter}\ and\ \citenamefont
  {Schulenburg}(2010)}]{Richter2010}%
  \BibitemOpen
  \bibfield  {author} {\bibinfo {author} {\bibfnamefont {J.}~\bibnamefont
  {Richter}}\ and\ \bibinfo {author} {\bibfnamefont {J.}~\bibnamefont
  {Schulenburg}},\ }\bibfield  {title} {\enquote {\bibinfo {title} {The
  {{Spin}}-1/2 $J_1-J_2$ {{Heisenberg
  Antiferromagnet}} on the {{Square Lattice}}: {{Exact Diagonalization}} for
  {{N}}=40 {{Spins}}},}\ }\href {\doibase 10.1140/epjb/e2009-00400-4}
  {\bibfield  {journal} {\bibinfo  {journal} {The European Physical Journal B}\
  }\textbf {\bibinfo {volume} {73}},\ \bibinfo {pages} {117--124} (\bibinfo
  {year} {2010})}\BibitemShut {NoStop}%
\bibitem [{\citenamefont {Viana}\ \emph {et~al.}(2017)\citenamefont {Viana},
  \citenamefont {Rodriguez~Salmon}, \citenamefont {Neto},\ and\ \citenamefont
  {Carvalho}}]{Viana2017}%
  \BibitemOpen
  \bibfield  {author} {\bibinfo {author} {\bibfnamefont {Roberto~J.}\
  \bibnamefont {Viana}}, \bibinfo {author} {\bibfnamefont {Octavio~D.}\
  \bibnamefont {Rodriguez~Salmon}}, \bibinfo {author} {\bibfnamefont
  {Minos~A.}\ \bibnamefont {Neto}}, \ and\ \bibinfo {author} {\bibfnamefont
  {Diego~C.}\ \bibnamefont {Carvalho}},\ }\bibfield  {title} {\enquote
  {\bibinfo {title} {A new effective correlation mean-field theory for the
  ferromagnetic spin-1 {{Blume}}\textendash{}{{Capel}} model in a transverse
  crystal field},}\ }\href {\doibase 10.1142/S0217979218500388} {\bibfield
  {journal} {\bibinfo  {journal} {International Journal of Modern Physics B}\
  }\textbf {\bibinfo {volume} {32}},\ \bibinfo {pages} {1850038} (\bibinfo
  {year} {2017})}\BibitemShut {NoStop}%
\bibitem [{\citenamefont {Santos}\ \emph {et~al.}(2018)\citenamefont {Santos},
  \citenamefont {{da Costa}},\ and\ \citenamefont {{de
  Ara{\'u}jo}}}]{Santos2018}%
  \BibitemOpen
  \bibfield  {author} {\bibinfo {author} {\bibfnamefont {P.~V.}\ \bibnamefont
  {Santos}}, \bibinfo {author} {\bibfnamefont {F.~A.}\ \bibnamefont {{da
  Costa}}}, \ and\ \bibinfo {author} {\bibfnamefont {J.~M.}\ \bibnamefont {{de
  Ara{\'u}jo}}},\ }\bibfield  {title} {\enquote {\bibinfo {title} {The {{Random
  Field Blume}}-{{Capel Model Revisited}}},}\ }\href {\doibase
  10.1016/j.jmmm.2017.12.008} {\bibfield  {journal} {\bibinfo  {journal}
  {Journal of Magnetism and Magnetic Materials}\ }\textbf {\bibinfo {volume}
  {451}},\ \bibinfo {pages} {737--740} (\bibinfo {year} {2018})}\BibitemShut
  {NoStop}%
\bibitem [{\citenamefont {Blume}(1966)}]{Blume1966}%
  \BibitemOpen
  \bibfield  {author} {\bibinfo {author} {\bibfnamefont {M.}~\bibnamefont
  {Blume}},\ }\bibfield  {title} {\enquote {\bibinfo {title} {Theory of the
  {{First}}-{{Order Magnetic Phase Change}} in
  {{UO}}$_2$},}
  }\href {\doibase 10.1103/PhysRev.141.517} {\bibfield  {journal} {\bibinfo
  {journal} {Physical Review}\ }\textbf {\bibinfo {volume} {141}},\ \bibinfo
  {pages} {517--524} (\bibinfo {year} {1966})}\BibitemShut {NoStop}%
\bibitem [{\citenamefont {Capel}(1966)}]{Capel1966}%
  \BibitemOpen
  \bibfield  {author} {\bibinfo {author} {\bibfnamefont {H.~W.}\ \bibnamefont
  {Capel}},\ }\bibfield  {title} {\enquote {\bibinfo {title} {On the
  possibility of first-order phase transitions in {{Ising}} systems of triplet
  ions with zero-field splitting},}\ }\href {\doibase
  10.1016/0031-8914(66)90027-9} {\bibfield  {journal} {\bibinfo  {journal}
  {Physica}\ }\textbf {\bibinfo {volume} {32}},\ \bibinfo {pages} {966--988}
  (\bibinfo {year} {1966})}\BibitemShut {NoStop}%
\bibitem [{\citenamefont {Balcerzak}\ and\ \citenamefont
  {Tucker}(1995)}]{Balcerzak1995}%
  \BibitemOpen
  \bibfield  {author} {\bibinfo {author} {\bibfnamefont {T.}~\bibnamefont
  {Balcerzak}}\ and\ \bibinfo {author} {\bibfnamefont {J.~W.}\ \bibnamefont
  {Tucker}},\ }\bibfield  {title} {\enquote {\bibinfo {title} {Phase
  {{Transitions}} of a {{Spin}}-{{One Magnetic Film}}},}\ }\href {\doibase
  10.1016/0304-8853(94)00965-1} {\bibfield  {journal} {\bibinfo  {journal}
  {Journal of Magnetism and Magnetic Materials}\ }\textbf {\bibinfo {volume}
  {140-144}},\ \bibinfo {pages} {653--654} (\bibinfo {year}
  {1995})}\BibitemShut {NoStop}%
\bibitem [{\citenamefont {Fittipaldi}\ and\ \citenamefont
  {Kaneyoshi}(1989)}]{Fittipaldi1989}%
  \BibitemOpen
  \bibfield  {author} {\bibinfo {author} {\bibfnamefont {I.~P.}\ \bibnamefont
  {Fittipaldi}}\ and\ \bibinfo {author} {\bibfnamefont {T.}~\bibnamefont
  {Kaneyoshi}},\ }\bibfield  {title} {\enquote {\bibinfo {title} {Re-{{Entrant
  Behaviour}} of the {{Anisotropic BEG Model}} in the {{Effective}}-{{Field
  Approximation}}},}\ }\href {\doibase 10.1088/0953-8984/1/37/001} {\bibfield
  {journal} {\bibinfo  {journal} {Journal of Physics: Condensed Matter}\
  }\textbf {\bibinfo {volume} {1}},\ \bibinfo {pages} {6513} (\bibinfo {year}
  {1989})}\BibitemShut {NoStop}%
\bibitem [{\citenamefont {Wilding}\ and\ \citenamefont
  {Nielaba}(1996)}]{Wilding1996}%
  \BibitemOpen
  \bibfield  {author} {\bibinfo {author} {\bibfnamefont {N.~B.}\ \bibnamefont
  {Wilding}}\ and\ \bibinfo {author} {\bibfnamefont {P.}~\bibnamefont
  {Nielaba}},\ }\bibfield  {title} {\enquote {\bibinfo {title} {Tricritical
  universality in a two-dimensional spin fluid},}\ }\href {\doibase
  10.1103/PhysRevE.53.926} {\bibfield  {journal} {\bibinfo  {journal} {Phys.
  Rev. E}\ }\textbf {\bibinfo {volume} {53}},\ \bibinfo {pages} {926--934}
  (\bibinfo {year} {1996})}\BibitemShut {NoStop}%
\bibitem [{\citenamefont {Xavier}\ \emph {et~al.}(1998)\citenamefont {Xavier},
  \citenamefont {Alcaraz}, \citenamefont {Lara},\ and\ \citenamefont
  {Plascak}}]{Xavier1998}%
  \BibitemOpen
  \bibfield  {author} {\bibinfo {author} {\bibfnamefont {J.~C.}\ \bibnamefont
  {Xavier}}, \bibinfo {author} {\bibfnamefont {F.~C.}\ \bibnamefont {Alcaraz}},
  \bibinfo {author} {\bibfnamefont {D.~Pen{\~a}}\ \bibnamefont {Lara}}, \ and\
  \bibinfo {author} {\bibfnamefont {J.~A.}\ \bibnamefont {Plascak}},\
  }\bibfield  {title} {\enquote {\bibinfo {title} {Critical {{Behavior}} of the
  {{Spin}}-3/2 {{Blume}}-{{Capel Model}} in {{Two
  Dimensions}}},}\ }\href {\doibase 10.1103/PhysRevB.57.11575} {\bibfield
  {journal} {\bibinfo  {journal} {Physical Review B}\ }\textbf {\bibinfo
  {volume} {57}},\ \bibinfo {pages} {11575--11581} (\bibinfo {year}
  {1998})}\BibitemShut {NoStop}%
\bibitem [{\citenamefont {Silva}\ \emph {et~al.}(2006)\citenamefont {Silva},
  \citenamefont {Caparica},\ and\ \citenamefont {Plascak}}]{Silva2006}%
  \BibitemOpen
  \bibfield  {author} {\bibinfo {author} {\bibfnamefont {C.~J.}\ \bibnamefont
  {Silva}}, \bibinfo {author} {\bibfnamefont {A.~A.}\ \bibnamefont {Caparica}},
  \ and\ \bibinfo {author} {\bibfnamefont {J.~A.}\ \bibnamefont {Plascak}},\
  }\bibfield  {title} {\enquote {\bibinfo {title} {Wang-{{Landau Monte Carlo
  Simulation}} of the {{Blume}}-{{Capel Model}}},}\ }\href {\doibase
  10.1103/PhysRevE.73.036702} {\bibfield  {journal} {\bibinfo  {journal}
  {Physical Review E}\ }\textbf {\bibinfo {volume} {73}},\ \bibinfo {pages}
  {036702} (\bibinfo {year} {2006})}\BibitemShut {NoStop}%
\bibitem [{\citenamefont {Y{\"u}ksel}\ \emph {et~al.}(2009)\citenamefont
  {Y{\"u}ksel}, \citenamefont {Ak{\i}nc{\i}},\ and\ \citenamefont
  {Polat}}]{Yuksel2009}%
  \BibitemOpen
  \bibfield  {author} {\bibinfo {author} {\bibfnamefont {Yusuf}\ \bibnamefont
  {Y{\"u}ksel}}, \bibinfo {author} {\bibfnamefont {{\"U}mit}\ \bibnamefont
  {Ak{\i}nc{\i}}}, \ and\ \bibinfo {author} {\bibfnamefont {Hamza}\
  \bibnamefont {Polat}},\ }\bibfield  {title} {\enquote {\bibinfo {title} {An
  {{Introduced Effective}}-{{Field Approximation}} and {{Monte Carlo Study}} of
  a {{Spin}}-1 {{Blume}}\textendash{}{{Capel Model}} on a {{Square
  Lattice}}},}\ }\href {\doibase 10.1088/0031-8949/79/04/045009} {\bibfield
  {journal} {\bibinfo  {journal} {Physica Scripta}\ }\textbf {\bibinfo {volume}
  {79}},\ \bibinfo {pages} {045009} (\bibinfo {year} {2009})}\BibitemShut
  {NoStop}%
\bibitem [{\citenamefont {Malakis}\ \emph {et~al.}(2010)\citenamefont
  {Malakis}, \citenamefont {Berker}, \citenamefont {Hadjiagapiou},
  \citenamefont {Fytas},\ and\ \citenamefont {Papakonstantinou}}]{Malakis2010}%
  \BibitemOpen
  \bibfield  {author} {\bibinfo {author} {\bibfnamefont {A.}~\bibnamefont
  {Malakis}}, \bibinfo {author} {\bibfnamefont {A.~Nihat}\ \bibnamefont
  {Berker}}, \bibinfo {author} {\bibfnamefont {I.~A.}\ \bibnamefont
  {Hadjiagapiou}}, \bibinfo {author} {\bibfnamefont {N.~G.}\ \bibnamefont
  {Fytas}}, \ and\ \bibinfo {author} {\bibfnamefont {T.}~\bibnamefont
  {Papakonstantinou}},\ }\bibfield  {title} {\enquote {\bibinfo {title}
  {Multicritical points and crossover mediating the strong violation of
  universality: Wang-landau determinations in the random-bond $d=2$ Blume-Capel
  model},}\ }\href {\doibase 10.1103/PhysRevE.81.041113} {\bibfield  {journal}
  {\bibinfo  {journal} {Phys. Rev. E}\ }\textbf {\bibinfo {volume} {81}},\
  \bibinfo {pages} {041113} (\bibinfo {year} {2010})}\BibitemShut {NoStop}%
\bibitem [{\citenamefont {Plascak}\ and\ \citenamefont
  {Martins}(2013)}]{Plascak2013}%
  \BibitemOpen
  \bibfield  {author} {\bibinfo {author} {\bibfnamefont {J.A.}\ \bibnamefont
  {Plascak}}\ and\ \bibinfo {author} {\bibfnamefont {P.H.L.}\ \bibnamefont
  {Martins}},\ }\bibfield  {title} {\enquote {\bibinfo {title} {Probability
  distribution function of the order parameter: Mixing fields and
  universality},}\ }\href {\doibase https://doi.org/10.1016/j.cpc.2012.09.014}
  {\bibfield  {journal} {\bibinfo  {journal} {Computer Physics Communications}\
  }\textbf {\bibinfo {volume} {184}},\ \bibinfo {pages} {259 -- 269} (\bibinfo
  {year} {2013})}\BibitemShut {NoStop}%
  \bibitem [{\citenamefont {Ryu}\ and\ \citenamefont {Kwak}(2013)}]{Ryu2013}%
  \BibitemOpen
  \bibfield  {author} {\bibinfo {author} {\bibfnamefont {Seol}\ \bibnamefont
  {Ryu}}\ and\ \bibinfo {author} {\bibfnamefont {Wooseop}\ \bibnamefont
  {Kwak}},\ }\bibfield  {title} {{\enquote {\bibinfo
  {title} {An efficient determination of the critical temperature of the
  {Blume}-{Capel} model},}\ }}\href {\doibase 10.3938/jkps.62.861} {\bibfield
  {journal} {\bibinfo  {journal} {Journal of the Korean Physical Society}\
  }\textbf {\bibinfo {volume} {62}},\ \bibinfo {pages} {861--865} (\bibinfo
  {year} {2013})}\BibitemShut {NoStop}%
\bibitem [{\citenamefont {Kwak}\ \emph {et~al.}(2015)\citenamefont {Kwak},
  \citenamefont {Jeong}, \citenamefont {Lee},\ and\ \citenamefont
  {Kim}}]{Kwak2015}%
  \BibitemOpen
  \bibfield  {author} {\bibinfo {author} {\bibfnamefont {Wooseop}\ \bibnamefont
  {Kwak}}, \bibinfo {author} {\bibfnamefont {Joohyeok}\ \bibnamefont {Jeong}},
  \bibinfo {author} {\bibfnamefont {Juhee}\ \bibnamefont {Lee}}, \ and\
  \bibinfo {author} {\bibfnamefont {Dong-Hee}\ \bibnamefont {Kim}},\ }\bibfield
   {title} {\enquote {\bibinfo {title} {First-order phase transition and
  tricritical scaling behavior of the {Blume}-{Capel} model: {A}
  {Wang}-{Landau} sampling approach},}\ }\href {\doibase
  10.1103/PhysRevE.92.022134} {\bibfield  {journal} {\bibinfo  {journal}
  {Physical Review E}\ }\textbf {\bibinfo {volume} {92}},\ \bibinfo {pages}
  {022134} (\bibinfo {year} {2015})}\BibitemShut {NoStop}%
\bibitem [{\citenamefont {Zierenberg}\ \emph {et~al.}(2017)\citenamefont
  {Zierenberg}, \citenamefont {Fytas}, \citenamefont {Weigel}, \citenamefont
  {Janke},\ and\ \citenamefont {Malakis}}]{Zierenberg2017}%
  \BibitemOpen
  \bibfield  {author} {\bibinfo {author} {\bibfnamefont {Johannes}\
  \bibnamefont {Zierenberg}}, \bibinfo {author} {\bibfnamefont {Nikolaos~G.}\
  \bibnamefont {Fytas}}, \bibinfo {author} {\bibfnamefont {Martin}\
  \bibnamefont {Weigel}}, \bibinfo {author} {\bibfnamefont {Wolfhard}\
  \bibnamefont {Janke}}, \ and\ \bibinfo {author} {\bibfnamefont {Anastasios}\
  \bibnamefont {Malakis}},\ }\bibfield  {title} {{\enquote
  {\bibinfo {title} {Scaling and universality in the phase diagram of the {2D}
  {Blume}-{Capel} model},}\ }}\href {\doibase 10.1140/epjst/e2016-60337-x}
  {\bibfield  {journal} {\bibinfo  {journal} {The European Physical Journal
  Special Topics}\ }\textbf {\bibinfo {volume} {226}},\ \bibinfo {pages}
  {789--804} (\bibinfo {year} {2017})}\BibitemShut {NoStop}%
\bibitem [{\citenamefont {Viana}\ \emph {et~al.}(2016)\citenamefont {Viana},
  \citenamefont {Salmon},\ and\ \citenamefont {Neto}}]{Viana2016}%
  \BibitemOpen
  \bibfield  {author} {\bibinfo {author} {\bibfnamefont {J.~Roberto}\
  \bibnamefont {Viana}}, \bibinfo {author} {\bibfnamefont {Octavio
  D.~Rodriguez}\ \bibnamefont {Salmon}}, \ and\ \bibinfo {author}
  {\bibfnamefont {Minos~A.}\ \bibnamefont {Neto}},\ }\bibfield  {title}
  {\enquote {\bibinfo {title} {A {{Mean}}-{{Field Approach Applied}} for the
  {{Ferromagnetic Spin}}-1 {{Blume}}-{{Capel Model}}},}\ }\href@noop {}
  {\bibfield  {journal} {\bibinfo  {journal} {arXiv:1607.01270 [cond-mat]}\ }
  (\bibinfo {year} {2016})},\ \Eprint {http://arxiv.org/abs/1607.01270}
  {arXiv:1607.01270} \BibitemShut {NoStop}%
\bibitem [{\citenamefont {Butera}\ and\ \citenamefont
  {Pernici}(2018)}]{Butera2018}%
  \BibitemOpen
  \bibfield  {author} {\bibinfo {author} {\bibfnamefont {P.}~\bibnamefont
  {Butera}}\ and\ \bibinfo {author} {\bibfnamefont {M.}~\bibnamefont
  {Pernici}},\ }\bibfield  {title} {\enquote {\bibinfo {title} {The
  {Blume}–{Capel} model for spins {S}=1 and 3/2 in dimensions d=2 and 3},}\
  }\href {\doibase 10.1016/j.physa.2018.05.010} {\bibfield  {journal} {\bibinfo
   {journal} {Physica A: Statistical Mechanics and its Applications}\ }\textbf
  {\bibinfo {volume} {507}},\ \bibinfo {pages} {22--66} (\bibinfo {year}
  {2018})}\BibitemShut {NoStop}%
\bibitem [{\citenamefont {Jung}\ and\ \citenamefont {Kim}(2017)}]{Jung2017}%
  \BibitemOpen
  \bibfield  {author} {\bibinfo {author} {\bibfnamefont {Moonjung}\
  \bibnamefont {Jung}}\ and\ \bibinfo {author} {\bibfnamefont {Dong-Hee}\
  \bibnamefont {Kim}},\ }\bibfield  {title} {\enquote {\bibinfo {title}
  {First-order transitions and thermodynamic properties in the 2d
  {Blume}-{Capel} model: the transfer-matrix method revisited},}\ }\href
  {\doibase 10.1140/epjb/e2017-80471-2} {\bibfield  {journal} {\bibinfo
  {journal} {The European Physical Journal B}\ }\textbf {\bibinfo {volume}
  {90}},\ \bibinfo {pages} {245} (\bibinfo {year} {2017})}\BibitemShut
  {NoStop}%
  \bibitem [{\citenamefont {Kim}\ and\ \citenamefont {Kwak}(2014)}]{Kim2014}%
  \BibitemOpen
  \bibfield  {author} {\bibinfo {author} {\bibfnamefont {Seung-Yeon}\
  \bibnamefont {Kim}}\ and\ \bibinfo {author} {\bibfnamefont {Wooseop}\
  \bibnamefont {Kwak}},\ }\bibfield  {title} {{\enquote
  {\bibinfo {title} {Study of the antiferromagnetic {Blume}-{Capel} model by
  using the partition function zeros in the complex temperature plane},}\
  }}\href {\doibase 10.3938/jkps.65.436} {\bibfield  {journal} {\bibinfo
  {journal} {Journal of the Korean Physical Society}\ }\textbf {\bibinfo
  {volume} {65}},\ \bibinfo {pages} {436--440} (\bibinfo {year}
  {2014})}\BibitemShut {NoStop}%
\bibitem [{\citenamefont {Badehdah}\ \emph {et~al.}(1998)\citenamefont
  {Badehdah}, \citenamefont {Bekhechi}, \citenamefont {Benyoussef},\ and\
  \citenamefont {Touzani}}]{Badehdah1998}%
  \BibitemOpen
  \bibfield  {author} {\bibinfo {author} {\bibfnamefont {M.}~\bibnamefont
  {Badehdah}}, \bibinfo {author} {\bibfnamefont {S.}~\bibnamefont {Bekhechi}},
  \bibinfo {author} {\bibfnamefont {A.}~\bibnamefont {Benyoussef}}, \ and\
  \bibinfo {author} {\bibfnamefont {M.}~\bibnamefont {Touzani}},\ }\bibfield
  {title} {{\enquote {\bibinfo {title} {Phase transition in
  the {{Blume}}-{{Capel}} model with second neighbour interaction},}\ }}\href
  {\doibase 10.1007/s100510050400} {\bibfield  {journal} {\bibinfo  {journal}
  {The European Physical Journal B - Condensed Matter and Complex Systems}\
  }\textbf {\bibinfo {volume} {4}},\ \bibinfo {pages} {431--440} (\bibinfo
  {year} {1998})}\BibitemShut {NoStop}%
\bibitem [{\citenamefont {Balcerzak}\ \emph {et~al.}(2014)\citenamefont
  {Balcerzak}, \citenamefont {Sza{\l}owski}, \citenamefont {Ja{\v s}{\v c}ur},
  \citenamefont {{\v Z}ukovi{\v c}}, \citenamefont {Bob{\'a}k},\ and\
  \citenamefont {Borovsk{\'y}}}]{Balcerzak2014}%
  \BibitemOpen
  \bibfield  {author} {\bibinfo {author} {\bibfnamefont {T.}~\bibnamefont
  {Balcerzak}}, \bibinfo {author} {\bibfnamefont {K.}~\bibnamefont
  {Sza{\l}owski}}, \bibinfo {author} {\bibfnamefont {M.}~\bibnamefont {Ja{\v
  s}{\v c}ur}}, \bibinfo {author} {\bibfnamefont {M.}~\bibnamefont {{\v
  Z}ukovi{\v c}}}, \bibinfo {author} {\bibfnamefont {A.}~\bibnamefont
  {Bob{\'a}k}}, \ and\ \bibinfo {author} {\bibfnamefont {M.}~\bibnamefont
  {Borovsk{\'y}}},\ }\bibfield  {title} {\enquote {\bibinfo {title}
  {Thermodynamic {{Description}} of the {{Ising Antiferromagnet}} on a
  {{Triangular Lattice}} with {{Selective Dilution}} by a {{Modified
  Pair}}-{{Approximation Method}}},}\ }\href {\doibase
  10.1103/PhysRevE.89.062140} {\bibfield  {journal} {\bibinfo  {journal}
  {Physical Review E}\ }\textbf {\bibinfo {volume} {89}},\ \bibinfo {pages}
  {062140} (\bibinfo {year} {2014})}\BibitemShut {NoStop}%
\bibitem [{\citenamefont {Balcerzak}(2004)}]{Balcerzak2004}%
  \BibitemOpen
  \bibfield  {author} {\bibinfo {author} {\bibfnamefont {T.}~\bibnamefont
  {Balcerzak}},\ }\bibfield  {title} {\enquote {\bibinfo {title} {Application
  of the {{Pair Approximation Method}} for the {{RKKY Interaction}}},}\ }\href
  {\doibase 10.1016/j.jmmm.2003.12.390} {\bibfield  {journal} {\bibinfo
  {journal} {Journal of Magnetism and Magnetic Materials}\ }\textbf {\bibinfo
  {volume} {272-276}},\ \bibinfo {pages} {e1035--e1036} (\bibinfo {year}
  {2004})}\BibitemShut {NoStop}%
\bibitem [{\citenamefont {Costabile}\ \emph {et~al.}(2012)\citenamefont
  {Costabile}, \citenamefont {Amazonas}, \citenamefont {Viana},\ and\
  \citenamefont {{de Sousa}}}]{Costabile2012}%
  \BibitemOpen
  \bibfield  {author} {\bibinfo {author} {\bibfnamefont {Emanuel}\ \bibnamefont
  {Costabile}}, \bibinfo {author} {\bibfnamefont {Marcio~A.}\ \bibnamefont
  {Amazonas}}, \bibinfo {author} {\bibfnamefont {J.~Roberto}\ \bibnamefont
  {Viana}}, \ and\ \bibinfo {author} {\bibfnamefont {J.~Ricardo}\ \bibnamefont
  {{de Sousa}}},\ }\bibfield  {title} {\enquote {\bibinfo {title} {Study of the
  {{First}}-{{Order Transition}} in the {{Spin}}-1
  {{Blume}}\textendash{}{{Capel Model}} by {{Using Effective}}-{{Field
  Theory}}},}\ }\href {\doibase 10.1016/j.physleta.2012.09.003} {\bibfield
  {journal} {\bibinfo  {journal} {Physics Letters A}\ }\textbf {\bibinfo
  {volume} {376}},\ \bibinfo {pages} {2922--2925} (\bibinfo {year}
  {2012})}\BibitemShut {NoStop}%
\bibitem [{\citenamefont {Balcerzak}(2008)}]{Balcerzak2008}%
  \BibitemOpen
  \bibfield  {author} {\bibinfo {author} {\bibfnamefont {T.}~\bibnamefont
  {Balcerzak}},\ }\bibfield  {title} {\enquote {\bibinfo {title}
  {Thermodynamics of the {{Small Clusters}}: {{The Gibbs Free}}-{{Energy
  Derivation}} for the {{Honmura}}\textendash{}{{Kaneyoshi Method}}},}\ }\href
  {\doibase 10.1016/j.jmmm.2008.05.015} {\bibfield  {journal} {\bibinfo
  {journal} {Journal of Magnetism and Magnetic Materials}\ }\textbf {\bibinfo
  {volume} {320}},\ \bibinfo {pages} {2359--2363} (\bibinfo {year}
  {2008})}\BibitemShut {NoStop}%
\bibitem [{\citenamefont {Peierls}(1938)}]{Peierls1938}%
  \BibitemOpen
  \bibfield  {author} {\bibinfo {author} {\bibfnamefont {R.}~\bibnamefont
  {Peierls}},\ }\bibfield  {title} {\enquote {\bibinfo {title} {On a {{Minimum
  Property}} of the {{Free Energy}}},}\ }\href {\doibase
  10.1103/PhysRev.54.918} {\bibfield  {journal} {\bibinfo  {journal} {Physical
  Review}\ }\textbf {\bibinfo {volume} {54}},\ \bibinfo {pages} {918--919}
  (\bibinfo {year} {1938})}\BibitemShut {NoStop}%
\bibitem [{\citenamefont {Feynman}(1955)}]{Feynman1955}%
  \BibitemOpen
  \bibfield  {author} {\bibinfo {author} {\bibfnamefont {R.~P.}\ \bibnamefont
  {Feynman}},\ }\bibfield  {title} {\enquote {\bibinfo {title} {Slow
  {{Electrons}} in a {{Polar Crystal}}},}\ }\href {\doibase
  10.1103/PhysRev.97.660} {\bibfield  {journal} {\bibinfo  {journal} {Physical
  Review}\ }\textbf {\bibinfo {volume} {97}},\ \bibinfo {pages} {660--665}
  (\bibinfo {year} {1955})}\BibitemShut {NoStop}%
\bibitem [{\citenamefont {Isihara}(1968)}]{Isihara1968}%
  \BibitemOpen
  \bibfield  {author} {\bibinfo {author} {\bibfnamefont {A.}~\bibnamefont
  {Isihara}},\ }\bibfield  {title} {\enquote {\bibinfo {title} {The
  {{Gibbs}}-{{Bogoliubov Inequality Dagger}}},}\ }\href {\doibase
  10.1088/0305-4470/1/5/305} {\bibfield  {journal} {\bibinfo  {journal}
  {Journal of Physics A: General Physics}\ }\textbf {\bibinfo {volume} {1}},\
  \bibinfo {pages} {539} (\bibinfo {year} {1968})}\BibitemShut {NoStop}%
\bibitem [{\citenamefont {Falk}(1970)}]{Falk1970}%
  \BibitemOpen
  \bibfield  {author} {\bibinfo {author} {\bibfnamefont {H.}~\bibnamefont
  {Falk}},\ }\bibfield  {title} {\enquote {\bibinfo {title} {Inequalities of
  {{J}}. {{W}}. {{Gibbs}}},}\ }\href {\doibase 10.1119/1.1976484} {\bibfield
  {journal} {\bibinfo  {journal} {American Journal of Physics}\ }\textbf
  {\bibinfo {volume} {38}},\ \bibinfo {pages} {858--869} (\bibinfo {year}
  {1970})}\BibitemShut {NoStop}%
\bibitem [{\citenamefont {Kuzemsky}(2015)}]{Kuzemsky2015}%
  \BibitemOpen
  \bibfield  {author} {\bibinfo {author} {\bibfnamefont {A.}~\bibfnamefont {L.}~\bibnamefont
  {Kuzemsky}},\ }\bibfield  {title} {\enquote {\bibinfo {title} {Variational principle of {B}ogoliubov and generalized mean fields in many-particle interacting systems},}\ }\href {\doibase 10.1142/S0217979215300108} {\bibfield
  {journal} {\bibinfo  {journal} {International Journal of Modern Physics B}\ }\textbf
  {\bibinfo {volume} {29}},\ \bibinfo {pages} {1530010} (\bibinfo {year}
  {2015})}\BibitemShut {NoStop}%
\bibitem [{\citenamefont {Yeomans}(1992)}]{Yeomans}%
  \BibitemOpen
  \bibfield  {author} {\bibinfo {author} {\bibfnamefont {J.}~{\bibfnamefont {M.}}~\bibnamefont
  {Yeomans}},\ }\bibfield  {title} {\enquote {\bibinfo {title} {Statistical {M}echanics of {P}hase {T}ransitions},}\ } {\bibinfo  {publisher} {Oxford University Press, Oxford}, }
  {\bibinfo {year}
  {1992}}\BibitemShut {NoStop}%
  \bibitem [{\citenamefont {Strecka}\ and\ \citenamefont
  {Jascur}(2015)}]{Strecka2015}%
  \BibitemOpen
  \bibfield  {author} {\bibinfo {author} {\bibfnamefont {J.}~\bibnamefont
  {Stre\v{c}ka}}\ and\ \bibinfo {author} {\bibfnamefont {M.}~\bibnamefont
  {Ja\v{s}\v{c}ur}},\ }\bibfield  {title} {\enquote {\bibinfo {title} {A brief account of the {I}sing and {I}sing-like models: {M}ean-field, effective-field and exact results},}\
  }\href {http://www.physics.sk/aps/pub.php?y=2015&pub=aps-15-04} {\bibfield  {journal} {\bibinfo
   {journal} {Acta Physica Slovaca}\ }\textbf
  {\bibinfo {volume} {65}},\ \bibinfo {pages} {235--367} (\bibinfo {year}
  {2015})}\BibitemShut {NoStop}%
\bibitem [{\citenamefont {Santos}(2017)}]{Santos2017}%
  \BibitemOpen
  \bibfield  {author} {\bibinfo {author} {\bibfnamefont {Jander~P.}\
  \bibnamefont {Santos}},\ }\bibfield  {title} {\enquote {\bibinfo {title} {A
  {{Generalization}} of {{Mean Field Theory}} in a {{Cluster}} with {{Many
  Sites}} on the {{Ising Model}} from the {{Bogoliubov Inequality}}:
  {{Hexagonal Nanowire}} and {{Nanotube}}},}\ }\href {\doibase
  10.1007/s13538-016-0478-4} {\bibfield  {journal} {\bibinfo  {journal}
  {Brazilian Journal of Physics}\ }\textbf {\bibinfo {volume} {47}},\ \bibinfo
  {pages} {122--130} (\bibinfo {year} {2017})}\BibitemShut {NoStop}%
\bibitem [{\citenamefont {Mendes}\ \emph {et~al.}(2018)\citenamefont {Mendes},
  \citenamefont {Barreto},\ and\ \citenamefont {Santos}}]{Mendes2018}%
  \BibitemOpen
  \bibfield  {author} {\bibinfo {author} {\bibfnamefont {R.~G.~B.}\
  \bibnamefont {Mendes}}, \bibinfo {author} {\bibfnamefont {F.~C.~S{\'a}}\
  \bibnamefont {Barreto}}, \ and\ \bibinfo {author} {\bibfnamefont {J.~P.}\
  \bibnamefont {Santos}},\ }\bibfield  {title} {\enquote {\bibinfo {title}
  {Thermodynamic {{States}} of the {{Mixed Spin}} 1/2 and {{Spin}} 1
  {{Hexagonal Nanowire System Obtained}} from a {{Seven}}-{{Site Cluster
  Within}} an {{Improved Mean Field Approximation}}},}\ }\href {\doibase
  10.1007/s13538-018-0560-1} {\bibfield  {journal} {\bibinfo  {journal}
  {Brazilian Journal of Physics}\ }\textbf {\bibinfo {volume} {48}},\ \bibinfo
  {pages} {137--145} (\bibinfo {year} {2018})}\BibitemShut {NoStop}%
\bibitem [{\citenamefont {S{\'a}~Barreto}\ and\ \citenamefont
  {Mota}(2012)}]{SaBarreto2012}%
  \BibitemOpen
  \bibfield  {author} {\bibinfo {author} {\bibfnamefont {F.~C.}\ \bibnamefont
  {S{\'a}~Barreto}}\ and\ \bibinfo {author} {\bibfnamefont {A.~L.}\
  \bibnamefont {Mota}},\ }\bibfield  {title} {\enquote {\bibinfo {title}
  {Correlations {{Equalities}} and {{Some Upper Bounds}} for the {{Critical
  Temperature}} for {{Spin One Systems}}},}\ }\href {\doibase
  10.1016/j.physa.2012.07.026} {\bibfield  {journal} {\bibinfo  {journal}
  {Physica A: Statistical Mechanics and its Applications}\ }\textbf {\bibinfo
  {volume} {391}},\ \bibinfo {pages} {5908--5917} (\bibinfo {year}
  {2012})}\BibitemShut {NoStop}%
\bibitem [{\citenamefont {Ak{\i}nc{\i}}(2012)}]{Akinci2012}%
  \BibitemOpen
  \bibfield  {author} {\bibinfo {author} {\bibfnamefont {{\"U}mit}\
  \bibnamefont {Ak{\i}nc{\i}}},\ }\bibfield  {title} {\enquote {\bibinfo
  {title} {An improved effective field theory formulation of spin-1 {{Ising}}
  systems with arbitrary coordination number z},}\ }\href@noop {} {\bibfield
  {journal} {\bibinfo  {journal} {arXiv:1201.3015 [cond-mat]}\ } (\bibinfo
  {year} {2012})},\ \Eprint {http://arxiv.org/abs/1201.3015} {arXiv:1201.3015
  [cond-mat]} \BibitemShut {NoStop}%
\bibitem [{\citenamefont {Enting}\ \emph {et~al.}(1994)\citenamefont {Enting},
  \citenamefont {Guttmann},\ and\ \citenamefont {Jensen}}]{Enting1994}%
  \BibitemOpen
  \bibfield  {author} {\bibinfo {author} {\bibfnamefont {I.~G.}\ \bibnamefont
  {Enting}}, \bibinfo {author} {\bibfnamefont {A.~J.}\ \bibnamefont
  {Guttmann}}, \ and\ \bibinfo {author} {\bibfnamefont {I.}~\bibnamefont
  {Jensen}},\ }\bibfield  {title} {\enquote {\bibinfo {title}
  {Low-{{Temperature Series Expansions}} for the {{Spin}}-1 {{Ising Model}}},}\
  }\href {\doibase 10.1088/0305-4470/27/21/014} {\bibfield  {journal} {\bibinfo
   {journal} {Journal of Physics A: Mathematical and General}\ }\textbf
  {\bibinfo {volume} {27}},\ \bibinfo {pages} {6987} (\bibinfo {year}
  {1994})}\BibitemShut {NoStop}%
\bibitem [{\citenamefont {Beale}(1986)}]{Beale1986}%
  \BibitemOpen
  \bibfield  {author} {\bibinfo {author} {\bibfnamefont {Paul~D.}\ \bibnamefont
  {Beale}},\ }\bibfield  {title} {\enquote {\bibinfo {title} {Finite-{{Size
  Scaling Study}} of the {{Two}}-{{Dimensional Blume}}-{{Capel Model}}},}\
  }\href {\doibase 10.1103/PhysRevB.33.1717} {\bibfield  {journal} {\bibinfo
  {journal} {Physical Review B}\ }\textbf {\bibinfo {volume} {33}},\ \bibinfo
  {pages} {1717--1720} (\bibinfo {year} {1986})}\BibitemShut {NoStop}%
  \bibitem [{\citenamefont {Schmidt}(2015)}]{Schmidt2015}%
  \BibitemOpen
  \bibfield  {author} {\bibinfo {author} {\bibfnamefont {M.}\ \bibnamefont
  {Schmidt}}, \bibinfo {author} {\bibfnamefont {F.~M.}\ \bibnamefont
  {Zimmer}}, \ and\ \bibinfo {author} {\bibfnamefont {S.~G.}~\bibnamefont
  {Magalhaes}},\ }\bibfield  {title} {\enquote {\bibinfo {title} {Weak randomness
in geometrically frustrated systems: {Spin}-glasses,},}\
  }\href {\doibase 10.1088/0031-8949/90/2/025809} {\bibfield  {journal} {\bibinfo
  {journal} {Physica Scripta}\ }\textbf {\bibinfo {volume} {90}},\ \bibinfo
  {pages} {025809} (\bibinfo {year} {2015})}\BibitemShut {NoStop}%
\end{thebibliography}

\end{document}